\documentclass[hyper,letterpaper]{JHEP3}
\usepackage{amsmath,amssymb,amsfonts,bm}
\usepackage{cite}
\usepackage{graphicx}
\usepackage{multirow}
\usepackage{verbatim}
\usepackage{appendix}

\usepackage{url}
\usepackage{float}

\newcommand{\bea}{\begin{eqnarray}}
\newcommand{\eea}{\end{eqnarray}}
\newcommand{\be}{\begin{equation}}
\newcommand{\ee}{\end{equation}}

\newcommand{\unknot}{{\,\raisebox{-.08cm}{\includegraphics[width=.37cm]{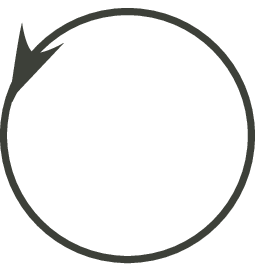}}\,}}

\newcommand{\R}{{\mathbb R}}
\newcommand{\C}{{\mathbb C}}

\newcommand{\cN}{{\mathcal{N}}}
\newcommand{\Li}{{\rm Li}}

\def\Tr{{\rm Tr \,}}

\newcommand{\cM}{{\cal M }}
\newcommand{\cC}{{\cal C }}

\newcommand{\cO}{{\cal O }}
\newcommand{\cH}{{\cal H }}

\renewcommand{\P}{{\cal P}}
\newcommand{\cp}{{\mathbb{C}}{\mathbf{P}}}
\renewcommand{\S}{{\bf S}}

\renewcommand{\bar}{\overline}
\renewcommand{\hat}{\widehat}

\title{Super-A-polynomial for knots and BPS states}

\author{Hiroyuki Fuji$^{1,2}$, Sergei Gukov$^{1,3}$ and Piotr Su{\l}kowski$^{1,4,5}$
\\
$^1$ California Institute of Technology, Pasadena, CA 91125, USA \\
$^2$ Nagoya University, Dept. of Physics, Graduate School of Science, \\
$\ $ Furo-cho, Chikusa-ku, Nagoya 464-8602, Japan \\
$^3$ Max-Planck-Institut f\"ur Mathematik, Vivatsgasse 7, D-53111 Bonn, Germany \\
$^4$ Institute for Theoretical Physics, University of Amsterdam, \\
$\ $ Science Park 904, 1090 GL, Amsterdam, The Netherlands \\
$^5$ Faculty of Physics, University of Warsaw, ul. Ho{\.z}a 69, 00-681 Warsaw, Poland}

\abstract{We introduce and compute a 2-parameter family deformation of the $A$-polynomial
that encodes the color dependence of the superpolynomial and that, in suitable limits,
reduces to various deformations of the $A$-polynomial studied in the literature.
These special limits include the $t$-deformation which leads to the ``refined $A$-polynomial''
introduced in the previous work of the authors
and the $Q$-deformation which leads, by the conjecture of Aganagic and Vafa,
to the augmentation polynomial of knot contact homology.
We also introduce and compute the quantum version of the super-$A$-polynomial,
an operator that encodes recursion relations for $S^r$-colored HOMFLY homology.
Much like its predecessor, the super-$A$-polynomial admits a simple physical interpretation
as the defining equation for the space of SUSY vacua (= critical points of the twisted superpotential)
in a circle compactification of the effective 3d $\cN=2$ theory associated to a knot or, more generally, to a 3-manifold $M$.
Equivalently, the algebraic curve defined by the zero locus of the super-$A$-polynomial
can be thought of as the space of open string moduli in a brane system associated with $M$.
As an inherent outcome of this work, we provide new interesting formulas for colored superpolynomials
for the trefoil and the figure-eight knot.
\\
\\
\\
\\
\\
\\
{\tt CALT-68-2870}}

\begin{document}


\section{Introduction and Summary}
\label{sec:intro}

The exact solution of $SL(2,\C)$ Chern-Simons theory with a Wilson loop
is determined by an algebraic curve \cite{Apol}:
\be
\cC: \quad \left\{~(x,y)\in\mathbb{C}^*\times \mathbb{C}^*~\Big|~
A(x,y) \; = \; 0 ~\right\}\,,
\label{Acurve}
\ee
namely the zero locus of the $A$-polynomial \cite{CCGL},
which plays a role similar to that of the Seiberg-Witten curve in $\cN=2$ gauge theory.
Various aspects of the $SL(2,\C)$ Chern-Simons partition function and its relation
to algebraic curves that appear in topological strings and supersymmetric gauge theories
were further studied in \cite{DijkgraafFuji-1,DGLZ,Wit-anal,Costantino,DGH,DijkgraafFuji-2,Tudor,TYsemiclass,DGSdual,abmodel,DGG}.

In particular, quantization of Chern-Simons theory turns a classical polynomial $A(x,y)$
into an operator of the form
\be
\hat A (\hat x, \hat y; q) \; = \; a_k \, \hat y^k + a_{k-1} \, \hat y^{k-1} + \ldots + a_1 \, \hat y + a_0 \,,
\label{Aform}
\ee
where $a_i \equiv a_i (\hat x, q)$ are rational functions of $\hat x$ and $q = e^{\hbar}$ (= the quantization parameter),
while $\hat x$ and $\hat y$ obey the commutation relation
\be
\hat y \hat x \; = \; q \hat x \hat y
\label{xycomm}
\ee
that follows directly from the symplectic structure on the classical phase space of Chern-Simons theory (see \cite{DGreview} for a review).
As a result, the algebraic curve \eqref{Acurve} that describes classical solutions in $SL(2,\C)$ Chern-Simons theory
upon quantization turns into a Schr\"odinger-like equation
\be
\hat A \; Z_{\text{CS}} \; = \; 0 \,,
\label{AonZ}
\ee
which leads to a set of recursion relations on polynomial invariants of the knot $K$.
Indeed, depending on whether the holonomy eigenvalues $x$ and $y$ take values in
the maximal torus of the group $G = SU(2)$ or its complexification $G_{\C} = SL(2,\C)$,
the same equation \eqref{AonZ} applies equally well to Chern-Simons theory
with the compact gauge group $SU(2)$ that computes the colored Jones polynomial $J_n (K;q)$
and to its analytic continuation that localizes on $SL(2,\C)$ flat connections.
In the former case, it leads to the so-called quantum volume conjecture~\cite{Apol}:
\be
\hat A \; J_* (K;q) \; = \; 0
\quad
\Leftrightarrow
\quad
a_k \, J_{n+k} (q) + \ldots + a_1 \, J_{n+1} (q) + a_0 \, J_n (q) \; = \; 0
\label{VCquant}
\ee
which in the mathematical literature was independently proposed around the same time~\cite{Garoufalidis} and is known as the AJ-conjecture.
The operators $\hat x$ and $\hat y$ act on the colored Jones polynomial as
\begin{align}
& \hat x J_n \; = \; q^n J_n \label{xyactionJ} \\
& \hat y J_n \; = \; J_{n+1}. \notag
\end{align}
In particular, one can easily verify that these operations obey the commutation relation \eqref{xycomm}
and that the Schr\"odinger-like equation \eqref{AonZ} with $\hat A (\hat x, \hat y; q)$ of the form \eqref{Aform}
is equivalent to the recursion relation \eqref{VCquant} that describes the ``color behavior'' of the colored Jones polynomial.

Besides the ``non-commutative'' deformation \eqref{Aform}-\eqref{xycomm},
the $A$-polynomial also admits two commutative deformations that 
in a similar way encode the ``color behavior'' of two natural generalizations of the colored Jones polynomial:
the $t$-deformation that corresponds to the categorification of colored Jones invariants \cite{FGS}
and $Q$-deformation that corresponds to extending $J_n (K;q)$ to higher rank knot polynomials \cite{AVqdef}.
A natural question is whether these two deformations can be combined in a single unifying structure?

The answer turns out to be ``yes'' and we call this unifying knot invariant the {\it super-$A$-polynomial} since it
describes how the $S^{n-1}$-colored superpolynomials $\P_n (a,q,t)$ depend on color, {\it i.e.} on the representation $R = S^{n-1}$,
much in the same way as $A$-polynomial does it for the colored Jones polynomial.\footnote{In fact,
in the context of open BPS invariants, the super-$A$-polynomial has already been computed and studied in \cite[sec. 4]{FGS}
for several prominent Calabi-Yau geometries.}
We remind that, in the context of BPS states, the superpolynomial
is defined as a generating function of refined open BPS invariants
on a rigid Calabi-Yau 3-fold $X$ in the presence of a Lagrangian brane supported on $L \subset X$:
\be
\P (a,q,t) \; := \; \Tr_{\cH^{\text{ref}}_{\text{BPS}}} \; a^{\beta} q^{P} t^{F} \,,
\qquad \beta \in H_2 (X,L)
\label{Paqt}
\ee
and, in application to knots, the superpolynomial $\P (K;a,q,t)$ is defined as a Poincar\'e polynomial
of the triply-graded homology theory $\cH (K)$ that categorifies the HOMFLY polynomial $P(K;a,q)$, see \cite{DGR} for details.
According to the conjecture of \cite{GSV}, these two definitions give the same result
when $X$ is the total space of the $\cO (-1) \oplus \cO (-1)$ bundle over $\cp^1$
and $L$ is the Lagrangian submanifold determined by the knot $K \subset \mathbf{S}^3$, {\it cf.} \cite{OoguriV,Taubes,Koshkin}.
Lagrangian branes of multiplicity $r=n-1$ yield the so-called ``$n$-colored'' version
of the superpolynomial which, in the context of knot homologies, was recently introduced in \cite{GS},
\be
\P_n (K; a,q,t) \; := \; \sum_{i,j,k} \, a^i q^j t^k \, \dim \cH^{S^{n-1}}_{i,j,k} (K) \,,
\label{superPdef}
\ee
as a Poincar\'e polynomial of a triply-graded homology theory categorifying
the $S^r$-colored HOMFLY polynomial (see also \cite{FGS,MMSS,ItoyamaMMM}).
\bigskip
\begin{figure}[ht]
\centering
\includegraphics[width=4.0in]{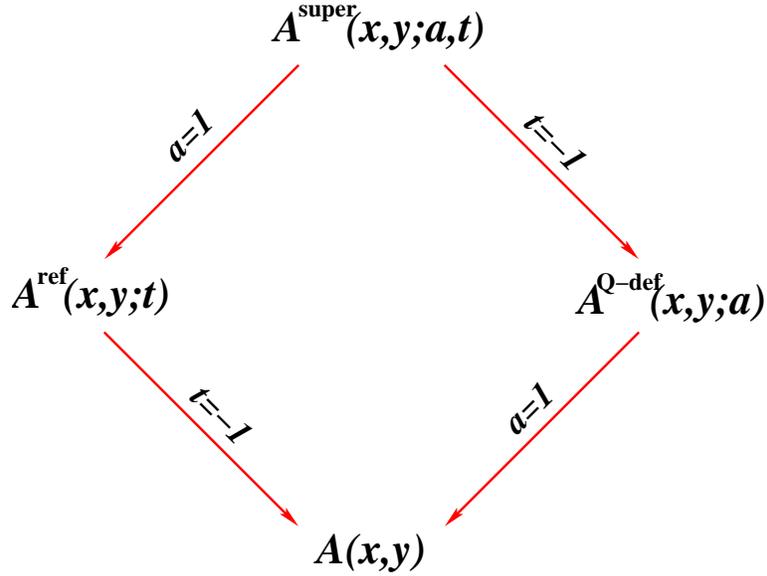}
\caption{Various specializations of the super-$A$-polynomial.}
\label{fig:superAlimits}
\end{figure}

The main goal of the present paper is to explain that $S^{n-1}$-colored superpolynomials $\P_n (K;a,q,t)$
depend on color ({\it i.e.} on the representation $R = S^{n-1}$) in a simple and controllable way,
governed by the super-$A$-polynomial $A^{\text{super}} (x,y;a,t)$
and by its quantization $\hat A^{\text{super}} (\hat x, \hat y;a,q,t)$.
Specifically, based on the physics arguments and the study of examples, we propose the following
analog of the generalized volume conjecture \cite{Apol} or its refined version~\cite{FGS}:

\medskip
\noindent
{\bf Conjecture 1:}
{\it In the limit}
\be
q = e^{\hbar} \to 1 \,, \qquad a = \text{fixed} \,, \qquad t = \text{fixed} \,, \qquad x = q^n = \text{fixed}
\label{reflimit}
\ee
{\it the $n$-colored superpolynomials $\P_n (K;a,q,t)$ exhibit the following ``large color'' behavior:}
\be
\P_n (K;a,q,t) \;\overset{{n \to \infty \atop \hbar \to 0}}{\sim}\;
\exp\left( \frac{1}{\hbar} \int \log y \frac{dx}{x} \,+\, \ldots \right)
\label{VCsuper}
\ee
{\it where ellipsis stand for regular terms (as $\hbar \to 0$) and the leading term is given by
the integral on the zero locus of the super-$A$-polynomial, {\it cf.} \eqref{Acurve}:}
\be
A^{\text{super}} (x,y;a,t) \; = \; 0 \,.
\label{supercurve}
\ee

Moreover, just like the ordinary $A$-polynomial has its quantum analog \eqref{Aform},
the super-$A$-polynomial is a characteristic polynomial of a quantum operator $\hat A^{\text{super}} (\hat x, \hat y;a,q,t)$
that combines commutative $t$- and $a$-deformations with the non-commutative $q$-deformation \eqref{xycomm}.
This ``quantum super-$A$-polynomial'' is the most advanced form of life in the world of $A$-polynomials
and knot homologies since it contains information about all deformations of the $A$-polynomial
and about all $n$-colored superpolynomials\footnote{that, in turn, contain information about colored $sl(N)$ knot homologies \cite{GS}}:

\medskip
\noindent
{\bf Conjecture 2:}
{\it For a given knot $K$, the colored superpolynomial $\P_n (K;a,q,t)$ satisfies a recurrence relation of the form \eqref{VCquant}:}
\be
a_k \, \P_{n+k} (K;a,q,t) + \ldots + a_1 \, \P_{n+1} (K;a,q,t) + a_0 \, \P_n (K;a,q,t) \; = \; 0
\label{QVCsuper}
\ee
{\it where $\hat x$ and $\hat y$ act on $\P_n (K;a,q,t)$ as in \eqref{xyactionJ},
and where the rational functions $a_i \equiv a_i (\hat x, a, q, t)$
are the coefficients of the ``quantum super-$A$-polynomial''}
\be
\hat A^{\text{super}} (\hat x, \hat y; a,q,t) \; = \; \sum_i a_i (\hat x, a, q, t) \, \hat y^i \,,
\label{Asuperform}
\ee
{\it whose characteristic polynomial is} $A^{\text{super}} (x,y;a,t)$.

As in \eqref{VCquant}, sometimes we informally write \eqref{QVCsuper} in the compact form
\be
\hat A^{\text{super}} \; \P_* (K;a,q,t) \; = \; 0 \,,   \label{AsuperP}
\ee
which is a quantum version of the classical curve \eqref{supercurve}.

\begin{table}[h]
\centering
\begin{tabular}{|@{$\Bigm|$}c|c@{$\Bigm|$}c|c@{$\Bigm|$}|}
\hline
\rule{0pt}{5mm}
\textbf{Quantum operator} & \textbf{provides recursion for} & \textbf{classical limit}  \\[3pt]
\hline
\hline
\rule{0pt}{5mm}
~~$\hat A^{\text{super}} (\hat x, \hat y; a,q,t)$~~ & ~~colored superpolynomial~~~ & ~~~$A^{\text{super}} (x,y;a,t)$~~  \\[3pt]
\hline
\rule{0pt}{5mm}
$\hat A^{\text{ref}} (\hat x, \hat y; q,t)$ & colored $sl(2)$ homology & $A^{\text{ref}} (x,y;t)$  \\[3pt]
\hline
\rule{0pt}{5mm}
$\hat A^{\text{Q-def}} (\hat x, \hat y; a,q)$ & colored HOMFLY & $A^{\text{Q-def}} (x,y;a)$  \\[3pt]
\hline
\rule{0pt}{5mm}
$\hat A (\hat x, \hat y; q)$ & colored Jones & $A (x,y)$  \\[3pt]
\hline
\end{tabular}
\caption{Quantum super-$A$-polynomial and its specializations lead to recursion relations for various $S^n$-colored knot invariants.\label{AAAtable}}
\end{table}

The superpolynomial unifies many polynomial and homological invariants of knots
that can be obtained from it via various specializations, applying differentials, {\it etc.}
For example, for $\cH$-thin knots the specialization to $a=q^2$ yields
the Poincar\'e polynomial of the colored $sl(2)$ knot homology.
Therefore, if $K$ is a thin knot ({\it e.g.} if $K$ is a two-bridge knot),
in the limit $a=q^2$ we expect \eqref{VCsuper} and \eqref{QVCsuper} to reproduce
the corresponding versions of the refined volume conjectures proposed in~\cite{FGS}.
In particular,
\be
\hat A^{\text{super}} (\hat x, \hat y; a=q^2,q,t) \; = \; \hat A^{\text{ref}} (\hat x, \hat y; q,t) \,,
\label{AsuperAref}
\ee
and, via further specialization to the classical limit $q=1$,
\be
A^{\text{super}} (x, y; a=1,t) \; = \; A^{\text{ref}} (x, y; t) \,.
\ee

Similarly, the specialization of the superpolynomial $\P_n (K;a,q,t)$
to $t=-1$ yields the HOMFLY polynomial or, in the problem at hand,
the {\it colored} HOMFLY polynomial~\cite{GS}. Therefore, at $t=-1$ the recursion relation \eqref{QVCsuper}
should reduce to the recursion relation for the $S^{n-1}$-colored HOMFLY polynomial,
whose characteristic variety --- called the $Q$-deformed $A$-polynomial in \cite{AVqdef} --- must be
contained in $A^{\text{super}} (x,y;a,t=-1)$ as a factor.
To avoid clutter, we include possible extra factors inherited\footnote{Some of these extra factors
will be explained below, in section \ref{sec:diff}.} from $A^{\text{super}} (x,y;a,t)$
in the definition of the $Q$-deformed $A$-polynomial, so that
\be
A^{\text{super}} (x, y; a,t = -1) \; = \; A^{\text{Q-def}} (x, y; a) \,.
\label{AsuperAaug}
\ee
Moreover, the authors of \cite{AVqdef} proposed an important conjecture that offers a new way of
looking at this polynomial (that, in our Figure \ref{fig:superAlimits}, occupies the right corner)
and identifies it with the augmentation polynomial of knot contact homology~\cite{NgFramed}.
Since we strongly believe this conjecture (and present some evidence for it below),
we are going to keep the notation $A^{\text{Q-def}} (x, y; a)$ but use the names ``$Q$-deformed $A$-polynomial''
and ``augmentation polynomial'' interchangeably in the rest of this paper (often we use both).
In fact, one justification for this comes from the fact (see \cite[Proposition 5.9]{NgFramed} for a proof)
that the classical augmentation polynomial, when specialized further to $a=1$,
reduces to the ordinary $A$-polynomial,
possibly with some extra factors, which altogether we denote simply by $A(x,y)$:
\be
A^{\text{super}} (x,y; a=1,t = -1) \; = \; A^{\text{Q-def}} (x,y; a=1) \; = \; A (x,y) \,,
\label{AugA}
\ee
as it should in order to fit perfectly in the diagram in Figure \ref{fig:superAlimits}.

Therefore, our super-$A$-polynomial $A^{\text{super}} (x,y;a,t)$ can be viewed,
on one hand, as a ``refinement'' of the augmentation polynomial $A^{\text{Q-def}} (x,y;a)$
and, on the other hand, as a ``$Q$-deformation'' of the refined $A$-polynomial $A^{\text{ref}} (x,y;t)$,
see Figure \ref{fig:superAlimits}.
Since both of these specializations have been recently computed for a number of simple knots \cite{FGS,ItoyamaMMM,AVqdef,NgFramed},
the time is just right for upgrading these results to the full-fledged super-$A$-polynomials, see table~\ref{table:superA}.

\begin{table}[h]
\centering
\begin{tabular}{|@{$\Bigm|$}c|@{$\Bigm|$}l|}
\hline
\textbf{Knot}  & $A^{\text{super}} (x,y;a,t)$  \\
\hline
\hline
Unknot, $\unknot$ & {\scriptsize $ ( - a^{-1} t^{-3})^{1/2} (1 + a t^3 x)  - (1 - x) y $}  \\
\hline
 & {\scriptsize $ a^2 t^5 (x-1)^2 x^2 + a t^2 x^2 (1 + a t^3 x)^2 y^3 +$}  \\
Figure-eight, ${\bf 4_1}$ & {\scriptsize $+ a t (x-1) (1 + t(1-t) x  + 2 a t^3(t+1) x^2
    -2 a t^4(t+1) x^3  + a^2 t^6(1-t) x^4  - a^2 t^8 x^5) y$} \\
    & {\scriptsize $- (1 + a t^3 x) (1 + a t(1-t) x +
    2 a t^2(t+1) x^2  + 2 a^2 t^4(t+1) x^3  + a^2 t^5(t-1) x^4  + a^3 t^7 x^5) y^2 $} \\
\hline
Trefoil, ${\bf 3_1}$ & {\scriptsize $a^2 t^4 (x-1) x^3 -a\big( 1 - t^2 x + 2 t^2 (1 + a t) x^2 + a t^5 x^3 + a^2 t^6 x^4 \big) y + (1 + a t^3 x) y^2$}  \\
\hline
$(2,2p$+$1)$ torus knot & {\scriptsize eliminate $z_0$ in   $ \left\{\begin{array}{l} 1 = \frac{(z_0-x)(t^2z_0-1)(1+ at^3 xz_0)}{t^{2+2p}z_0^{1+2p}(z_0-1)(atx+z_0)(t^2 x z_0-1)}   \\
y =  \frac{a^p t^{2 + 2 p} (x-1) x^{1 + 2 p} (atx + z_0) (1 + a t^3 x z_0)}{(1 + a t^3 x) (x - z_0) (t^2 x z_0-1)}   \end{array} \right.\  $ see tables \protect\ref{table_super}, \protect\ref{table_super4}, \protect\ref{table_super5}  }  \\
\hline
\end{tabular}
\caption{Super-$A$-polynomials for simple knots.\label{table:superA}}
\end{table}

Another interesting specialization of the super-$A$-polynomial involves
setting $x=1$
and provides a much more direct relation between the superpolynomial $\P
(a,q,t)$
and the super-$A$-polynomial of the same knot, {\it cf.}
table~\ref{table:superA}:
\be
A^{\text{super}} (x=1,y;a,t) \; = \; y^k + y^{k-1} \, \P (a,q=1,t) \,.
\label{superAsuperP}
\ee
In other words, it means that super-$A$-polynomials itself is not much
different from
the superpolynomial and contains information about the total dimension of
the HOMFLY homology
and also about the most interesting homological $t$-grading.
Further discussion of this property of the super-$A$-polynomial and related
aspects
of the colored HOMFLY homology will be discussed elsewhere.
Note, that specializing further to $a=1$ or $t=-1$, one can obtain similar
properties
of the polynomials $A^{\text{ref}} (x,y;t)$ or $A^{\text{Q-def}} (x,y;a)$
which, however, loose
some important information ({\it e.g.} in the specialization to $t=-1$ many
terms can cancel,
so that $A^{\text{Q-def}} (x=1,y;a)$ does not know about the total
dimension of the HOMFLY homology).


\section{Case studies}    \label{sec-case}

In this section we illustrate the ideas and the validity of the two Conjectures presented in the Introduction in explicit examples of various knots. We start with the simplest example of the unknot, and then discuss non-trivial examples of hyperbolic knots, such as figure-eight knot, and the entire family of $(2,2p+1)$ torus knots, with a special emphasis on the trefoil. In each case we start our considerations by providing explicit and general formulas for $S^{n-1}$-colored superpolynomials $\P_n (a,q,t)$, which are very interesting in their own right. In particular, we provide new expressions for colored superpolynomials for the trefoil and figure-eight knots, in a form which is particularly well suited to derive recursion relations which they satisfy and analyze their asymptotics. Taking advantage of these representations, subsequently we derive classical and quantum super-$A$-polynomials for these knots, discuss their properties and, in appropriate limits, relate them to other more familiar polynomials.

\subsection{Unknot}
\label{sec:unknot}

Let us start with the simplest example of the unknot, which we often denote as~$\unknot$. Despite its simplicity, this is still an interesting and important example; as we will see, some objects associated to the unknot, which are trivial in the non-refined and non-super case, become rather non-trivial when $t$- or $a$-dependence is turned on.

We recall than in the unknot case we must consider unreduced (or, sometimes also called ``unnormalized'') knot polynomials --
in particular, unreduced colored superpolynomial $\bar{\P}_{n}(a,q,t)$ -- since, by definition,
reduced polynomials are normalized by the value of the unknot, so that $\P_{n}(\unknot; a,q,t)=1$.
From the viewpoint of the (refined) Chern-Simons theory the unreduced
colored superpolynomial is defined as the ratio of partition functions
on $\S^3$ in the presence and absence of a knot. In case of the unknot
this ratio is given by the Macdonald polynomial, and after the change of
variables 
\be
a = A \left( \frac{q_1}{q_2} \right)^{3/2}, \qquad  q = \frac{1}{q_2}, \qquad   t = - \sqrt{\frac{q_2}{q_1}} \,, \label{GSvarchng}
\ee
we find that the $S^{n-1}$-colored superpolynomial $\bar{\P}_n (\unknot; a,q,t) \equiv \bar{\P}^{S^{n-1}} (\unknot; a,q,t)$ reads
\begin{eqnarray}
\bar{\P}_n (\unknot; a,q,t)
&=& \frac{Z_{SU(N)}^{\text{ref}}(\S^3,\unknot_{\Lambda^{n-1}};q_1,q_2)}{Z_{SU(N)}^{\text{ref}}(\S^3;q_1,q_2)}
=M_{\Lambda^{n-1}}(q_2^{\varrho};q_1,q_2) \nonumber \\
&=&
(-1)^{\frac{n-1}{2}}a^{-\frac{n-1}{2}}q^{\frac{n-1}{2}}t^{-\frac{3(n-1)}{2}} \frac{(-at^3;q)_{n-1}}{(q;q)_{n-1}} \,.
\label{Punknot}  
\end{eqnarray}
Note that, in our grading conventions, we need to consider Macdonald polynomials for anti-symmetric representations $\Lambda^{n-1}$,
$q_2^{\varrho}=(q_2^{\varrho_j})_{j=1,\ldots,N}$ with $\varrho_j=\frac{N+1}{2}-j$, and $a=q_2^{N-3/2} q_1^{3/2}$, see \cite{FGS} for more details.
We also use a standard notation for the $q$-Pochhammer symbol, which has the following asymptotics
\be
(x,q)_k = \prod_{i=0}^{k-1} (1 - x q^i) \sim e^{\frac{1}{\hbar}\left({\rm Li}_2(x)-{\rm Li}_2(x q^k)\right)}  . \label{qPoch}
\ee

\begin{figure}[ht]
\begin{center}
\includegraphics[width=0.6\textwidth]{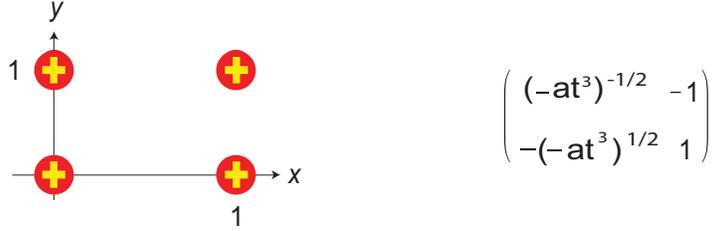}
\end{center}
\caption{Newton polygon for the super-$A$-polynomial of the unknot (left).
Red circles denote monomials of the super-$A$-polynomial, and smaller yellow crosses denote monomials of its $a=-t=1$ specialization.
In this example both Newton polygons look the same, so that positions of all circles and crosses overlap.
The coefficients of the super-$A$-polynomial are also shown in the matrix on the right.
The role of rows and columns is exchanged in these two presentations:
a monomial $a_{i,j}x^i y^j$ corresponds to a circle (resp. cross) at position $(i,j)$ in the Newton polygon,
while in the matrix on the right it is shown as the entry $a_{i,j}$ in the $(i+1)^{\text{th}}$ row and in the $(j+1)^{\text{th}}$ column.
These conventions are the same as in~\cite{FGS}.}
\label{fig-unknotNewtonMatrix}
\end{figure}

Once the general expression for the colored superpolynomial is found and presented in an appropriate form,
the next task is to find a recursion relation it satisfies.
In particular, as the homological unknot invariant (\ref{Punknot}) has a product form,
we can immediately write down the recursion relation it satisfies:
\be
\bar{\P}_{n+1} (\unknot; a,q,t) \; = \; (-a^{-1} t^{-3}q)^{1/2}  \frac{1 + a t^3 q^{n-1}}{1 - q^{n}}  \, \bar{\P}_n (\unknot; q,t) \,.
\label{Punrec}
\ee
This means  that the quantum super-$A$-polynomial for the unknot reads
\be
\hat A^{\text{super}}(\hat{x},\hat{y};a,q,t) \; = \;
( - a^{-1} t^{-3} q )^{1/2} (1 + a t^3 q^{-1} \hat{x})  - (1 - \hat x) \hat y \,.   \label{qAt-unknot}
\ee
In the classical limit $q \to 1$ this operator
reduces to the  classical super-$A$-polynomial defined by
\be
A^{\text{super}} (x,y;a,t) \; = \;
( - a^{-1} t^{-3})^{1/2} (1 + a t^3 x)  - (1 - x) y \,.   \label{At-unknot}
\ee
The Newton polygon as well as the coefficients of monomials of this polynomial are shown in figure~\ref{fig-unknotNewtonMatrix}.
On the other hand, in the unrefined limit $t=-1$ the relation (\ref{qAt-unknot}) takes the form
\be
\hat A^{\text{Q-def}}(\hat x, \hat y ;a,q) \; = \;
(a^{-1} q)^{1/2}(1 - a q^{-1} \hat x)  - (1 - \hat x) \hat y \,,
\ee
and specializing further to $q=1$ we get the augmentation polynomial
\be
A^{\textrm{Q-def}}(x,y;a) \; = \; a^{-1/2}(1 - a x)-(1-x)y \,.   \label{Aunknot1x1y}
\ee
Interestingly, this polynomial does not factorize, and only in the limit of ordinary $A$-polynomial $a\to 1$ do we get a factorized form with $y-1$ factor representing the abelian connection
\be
A(x,y) \; = \; (1 -  x)(1-y) \,.  \label{Axy-factor}
\ee

It is instructive to show that the super-$A$-polynomial (\ref{At-unknot}) can be also derived from the asymptotic analysis of~(\ref{Punknot}).
Indeed, using the asymptotics (\ref{qPoch}), in the limit (\ref{reflimit}) we can approximate (\ref{Punknot}) as
$$
P_n(\unknot; a,q,t) = \exp \frac{1}{\hbar}\Big(\log x\, \log(-a^{-1} t^{-3})^{1/2} + \textrm{Li}_2(x) - \textrm{Li}_2(-a t^3 x)  + \textrm{Li}_2(-a t^3) - \frac{\pi^2}{6} + \mathcal{O}(\hbar) \Big),
$$
from which identify the potential $\widetilde{\mathcal{W}} = \int \log y \frac{dx}{x}$ in \eqref{VCsuper} as
\be
\widetilde{\mathcal{W}} \; = \;  \log x\, \log(-a^{-1} t^{-3})^{1/2} + \textrm{Li}_2(x) - \textrm{Li}_2(-a t^3 x)  + \textrm{Li}_2(-a t^3) - \frac{\pi^2}{6} \,. \label{S0unknot}
\ee
Differentiating it with respect to $x$, we now obtain
\be
y = e^{x\partial_x \widetilde{\mathcal{W}} } = (-a^{-1} t^{-3})^{1/2} \frac{1 + a t^3 x}{1 - x} \,,
\ee
which reproduces the defining equation of the super-$A$-polynomial given in (\ref{At-unknot}).
We also note that for $a=-t=1$ the potential $\widetilde{\mathcal{W}}$ vanishes,
which is related to the factorization occurring in (\ref{Axy-factor}) and can be attributed
to the fact that the only $SL(2,\C)$ flat connections on a solid torus (= complement of the unknot) are abelian flat connections.
When $a\neq 1$ or $t\neq -1$, the potential $\widetilde{\mathcal{W}}$ is nonzero and presumably can be interpreted
as a contribution of ``deformed'' abelian flat connections. It would be interesting to pursue this interpretation further.


\subsection{Figure-eight knot}
\label{sec:fig8}

In this section we illustrate the program presented in the Introduction in the example of the figure-eight knot, also denoted ${\bf 4_1}$.
This is a hyperbolic knot, and we stress that it provides a highly non-trivial example,
for which many simplifications common in the realm of torus knots (to be discussed in the following sections) do not occur.

To start with, we present the colored superpolynomial (\ref{Paqt}) for this knot, denoted $\P_n ({\bf 4_1};a,q,t)$,
in the form most appropriate for finding its recursion relations~(\ref{QVCsuper}).
Then, we indeed find such recursion relations, thereby illustrating validity of the Conjecture~2
and, in the classical limit $q\to 1$, we find the form of the (classical) super-$A$-polynomial~(\ref{supercurve}).
We also derive the same super-$A$-polynomial from the analysis of the asymptotics (\ref{VCsuper}) of $\P_n (4_1;a,q,t)$,
thereby confirming the validity of the Conjecture~1 for this knot.
Finally, we discuss various properties of the classical and quantum super-$A$-polynomial
and show that in appropriate limits various familiar results are reproduced.

The form of the superpolynomial for the ${\bf 4_1}$ knot was recently proposed in \cite{ItoyamaMMM},
in certain variables and a grading choice which were natural from the viewpoint of the quantization rule proposed in that paper.
For our purposes it is more convenient to make a choice variables and grading conventions as in~\cite{FGS,GS}.
In appendix \ref{app-figure8} we summarize how the formula in \cite{ItoyamaMMM} was postulated and explain
how to transform it to the form convenient for our purposes.
Ultimately, after some rearrangements we find the following form of the colored superpolynomial for the ${\bf 4_1}$ knot
\be
\P_n ({\bf 4_1};a,q,t) = \sum_{k=0}^{\infty} (-1)^k a^{-k} t^{-2k} q^{-k(k-3)/2} \frac{(-a t q^{-1},q)_k}{(q,q)_k}  (q^{1-n},q)_k (-a t^3 q^{n-1}, q)_k \,.
\label{Paqt41}
\ee
Explicit values of $\P_n ({\bf 4_1};a,q,t)$ for low values of $n$ are given in table~\ref{tab-P41}.
The following checks confirm the validity of the expression (\ref{Paqt41}) in various special cases:
\begin{itemize}
\item for $a=q^2$ and $t=-1$, the above formula reduces to the familiar expression for the colored Jones polynomial studied {\it e.g.} in \cite{Habiro,Garoufalidis}:
$$
J_n ({\bf 4_1};q)=\P_n ({\bf 4_1};q^2,q,-1)= 
\sum_{k=0}^{n-1}  q^{n k} (q^{-n-1},q^{-1})_k (q^{-n+1},q)_k
$$
\item for $t=-1$ we checked that it agrees with the colored HOMFLY polynomial (\ref{PnKawagoe}) given in the unpublished work \cite{kawagoe}, which was also used in the analysis of \cite{AVqdef}; the precise relation is given in (\ref{Pour-Kawagoe});
\item for $n=2$ the superpolynomial (\ref{Paqt41}) agrees with the known result given {\it e.g.} in~\cite{DGR} (to match conventions we need to replace $a$ and $q$ in \cite{DGR} respectively by $a^{1/2}$ and $q^{1/2}$);
\item for $n=3$ and $n=4$ the expression (\ref{Paqt41}) reproduces results given in~\cite{GS};
\item for $a=-q^j t^k$ the expression \eqref{Paqt41} correctly reproduces specializations predicted from the colored / canceling differentials with $(a,q,t)$-grading $(-1,j,k)$, see~\cite{GS}.
\end{itemize}

\begin{table}[h]
\centering
\begin{tabular}{|@{$\Bigm|$}c|@{$\Bigm|$}l|}
\hline
$\, n$ & $\qquad \P_n ({\bf 4_1};a,q,t)$  \\
\hline
\hline
$\, 1$ & $1$   \\
\hline
$\, 2$ & $a^{-1}t^{-2} + t^{-1}q^{-1} + 1 + q t + a t^2$
\\
\hline
$\, 3$ & $ a^{-2} q^{-2} t^{-4} + (a^{-1} q^{-3} + a^{-1} q^{-2}) t^{-3} + (q^{-3} + a^{-1} q^{-1} + a^{-1}) t^{-2} +$ \\
& $+ (q^{-2} + q^{-1} + a^{-1} + a^{-1} q) t^{-1} + (q^{-1} + 3 + q) + (q^2 + q + a + a q^{-1}) t +$\\
& $+ (q^3 + a q + a) t^2 + (a q^3 + a q^2) t^3 + a^2 q^2 t^4 $\\
\hline
$\, 4$ & $1 + (1 + a^{-1} q t^{-1}) (1 + a^{-1} t^{-1}) (1 + a^{-1} q^{-1} t^{-1}) \times $\\
 & $\qquad \times   (1 + a^{-1} q^{-3} t^{-3}) (1 + a^{-1} q^{-4} t^{-3}) (1 + a^{-1} q^{-5} t^{-3}) a^3 q^6 t^6 +$\\
 & $+ (1 + q + q^2) (1 + a^{-1} q t^{-1}) (1 + a^{-1} q^{-3} t^{-3}) a t^2 +$ \\
 & $ + (1 + q + q^2) (1 + a^{-1} q t^{-1}) (1 + a^{-1} t^{-1}) (1 + a^{-1} q^{-3} t^{-3}) (1 + a^{-1} q^{-4} t^{-3}) a^2 q^2 t^4$\\
\hline
\end{tabular}
\caption{The colored superpolynomial of the ${\bf 4_1}$ knot for $n=1,2,3,4$.
\label{tab-P41} }
\end{table}

\begin{figure}[ht]
\begin{center}
\includegraphics[width=0.8\textwidth]{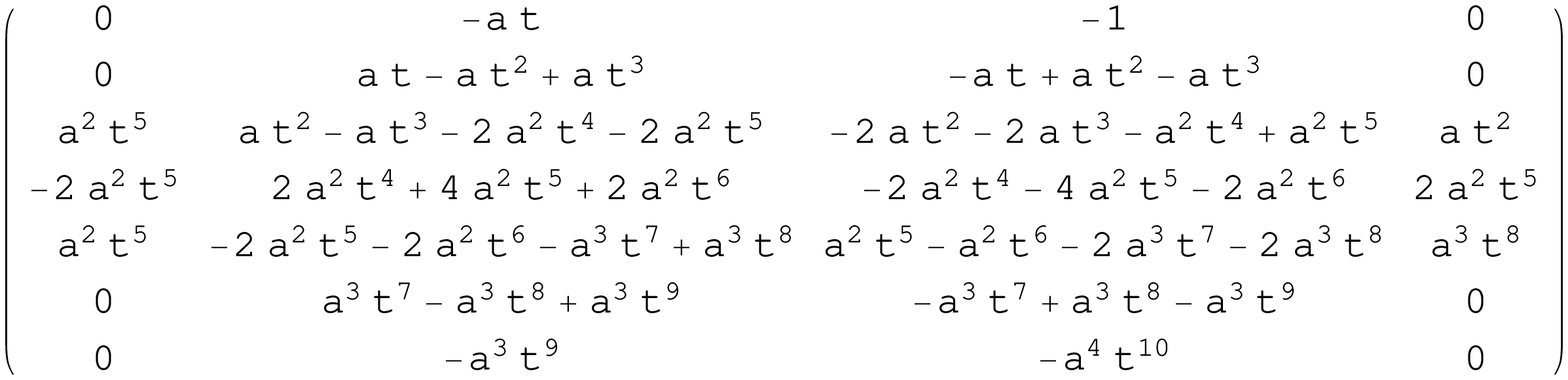}
\caption{Matrix form of the super-$A$-polynomial for the figure-eight knot. The conventions are the same as in the unknot example in figure \protect\ref{fig-unknotNewtonMatrix}.}
\label{fig:matrix41}
\end{center}
\end{figure}

\begin{figure}[ht]
\begin{center}
\includegraphics[width=0.5\textwidth]{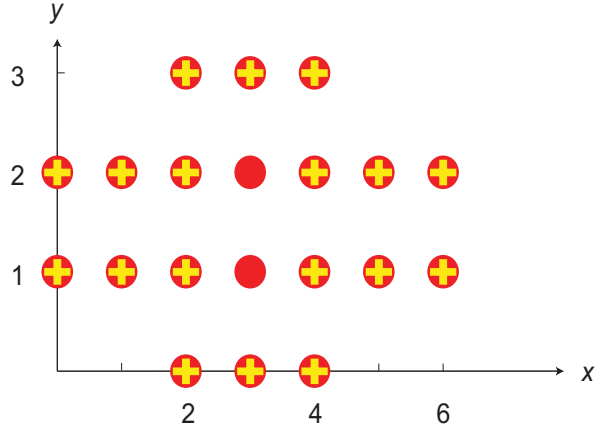}
\caption{Newton polygon of the super-$A$-polynomial for the figure-eight knot and its $a=-t=1$ limit. The conventions are the same as in figure \protect\ref{fig-unknotNewtonMatrix}.}
\label{fig:Newton41}
\end{center}
\end{figure}

In order to find a recursion relation satisfied by (\ref{Paqt41}) we use the Mathematica package \textrm{qZeil.m} developed by \cite{qZeil}.
With such a powerful tool, it is not hard to find the recursion, and in the notations of (\ref{Asuperform}) it takes the following four-term form
\be
\hat A^{\text{super}} (\hat x, \hat y; a,q,t) = a_0 + a_1 \hat{y} + a_2 \hat{y}^2 + a_3 \hat{y}^3,
\ee
where
\bea
a_0 & = &  \frac{a t^3 (1 - \hat{x}) (1 - q \hat{x}) (1 + a t^3 q^2 \hat{x}^2) (1 +
   a t^3 q^3 \hat{x}^2 )}{q^3 (1 + a t^3 \hat{x}) (1 + a  t^3 \hat{x}^2) (1 +
   a t^3 q \hat{x} ) (1 + a t^3 q^{-1} \hat{x}^2)}  \nonumber \\
a_1 & = & - \frac{(1 - q \hat{x}) (1 + a t^3 q^3\hat{x}^2 )}
 {t q^3 \hat{x}^2  (1 + a t^3 \hat{x}) (1 + a t^3 q \hat{x}) (1 + a t^3 q^{-1} \hat{x}^2)}   \nonumber \\
& & \quad \times \Big( 1 - t (t-1) q \hat{x} + a t^3 q^{-1}(1 + q^3 + q t + q^2 t) \hat{x}^2    \nonumber \\
 & & \qquad - a t^4 (q + q^2 + t + q^3 t)\hat{x}^3  - a^2 (t-1) t^6 q \hat{x}^4 - a^2 t^8 q^2 \hat{x}^5 \Big)   \nonumber \\
a_2 & = &  - \frac{(1 + a t^3 q^2 \hat{x}^2)}{a t^2 q^2 \hat{x}^2 (1 + a t^3 \hat{x}^2) (1 + a t^3 q \hat{x})}   \nonumber \\
& & \quad \times \Big( 1 - a t(t-1) \hat{x} + a  t^2 (q + q^2 + t + q^3 t) \hat{x}^2     \nonumber \\
 & & \qquad  + a^2  t^4 (1 + q^3 + q t + q^2 t) \hat{x}^3 + a^2 (t-1) t^5 q^3 \hat{x}^4 + a^3 t^7 q^3 \hat{x}^5  \Big)   \nonumber \\
a_3 & = &  1   \nonumber
\eea
Taking the classical limit $q\to 1$ (and clearing the denominators), we find the following classical super-$A$-polynomial
\bea
& & A^{\text{super}} (x, y; a,t) \, = \,
a^2 t^5 (x-1)^2 x^2 + a t^2 x^2 (1 + a t^3 x)^2 y^3 + \label{Asuper41} \\
& & \quad + a t (x-1) (1 + t(1-t) x  + 2 a t^3(t+1) x^2
    -2 a t^4(t+1) x^3  + a^2 t^6(1-t) x^4  - a^2 t^8 x^5) y  \nonumber \\
 & & \quad   - (1 + a t^3 x) (1 + a t(1-t) x +
    2 a t^2(t+1) x^2  + 2 a^2 t^4(t+1) x^3  + a^2 t^5(t-1) x^4  + a^3 t^7 x^5) y^2 . \nonumber
\eea
The coefficients of the monomials in this polynomial are assembled into a matrix form presented in figure~\ref{fig:matrix41},
and the corresponding Newton polygon is given in figure~\ref{fig:Newton41}.

According to the Conjecture~1, we should be able to reproduce the same polynomial from the asymptotic behavior of the colored superpolynomial~(\ref{Paqt41}).
This is indeed the case. To show this, we introduce the variable $z=e^{\hbar k}$. Then, in the limit (\ref{reflimit}) with $z=\text{const}$
the sum over $k$ in (\ref{Paqt41}) can be approximated by the integral
\be
\P_n ({\bf 4_1};a,q,t) \; \sim \; \int dz\; e^{\frac{1}{\hbar}\left(\widetilde{\mathcal{W}}({\bf 4_1};z,x)+{\cal O}(\hbar)\right)} \,.   \label{Pn41-integral}
\ee
The potential $\widetilde{\mathcal{W}}({\bf 4_1};z,x)$ can be determined from the asymptotics (\ref{qPoch}):
\bea
& & \widetilde{\mathcal{W}}({\bf 4_1};z,x)  =  \pi i \log z - \frac{\pi^2}{6}- (\log a + 2\log t) \log z - \frac{1}{2} (\log z)^2   \label{V41}\\
& & \quad   + \Li_2( x^{-1}) - \Li_2(x^{-1}z) + \Li_2(-a t) - \Li_2(-a t z) + \Li_2(-a x t^3)
-  \Li_2(-a x t^3 z)  - \Li_2(z) \,.  \nonumber
\eea
At the saddle point
\be
\frac{\partial \widetilde{\mathcal{W}}({\bf 4_1};z,x)}{\partial z}\Bigg|_{z=z_0}=0
\label{saddle_point41}
\ee
it determines the leading asymptotic behavior (\ref{VCsuper}), which at the same time is also computed by the integral along the curve~(\ref{supercurve}),
implying a key identity
\be
y = \exp\left(x\frac{\partial \widetilde{\mathcal{W}}({\bf 4_1};z_0,x)}{\partial x}\right) \,.    \label{yV}
\ee
Plugging the expression (\ref{V41}) to the above two equations we obtain the following system
\be
\left\{\begin{array}{l} 1 = \frac{(x - z_0) (1 + a t z_0) (1 + a t^3 x z_0)}{a t^2 x z_0 (z_0-1)}   \\
y =  \frac{(x-1) (1 + a t^3 x z_0)}{(1 + a t^3 x) (x - z_0)}   \end{array} \right.
\ee
Eliminating $z_0$ from these two equations we indeed reproduce the super-$A$-polynomial~(\ref{Asuper41}).
Overall, the above statements verify the validity of the Conjecture~1 and Conjecture~2 for the figure-eight knot.

The relation (\ref{superAsuperP}) can be checked explicitly from
(\ref{Paqt41}) and (\ref{Asuper41}). The colored super-A-polynomial 
$\P_{n=2}({\bf 4_1};a,q,t)$ reduces under $q=1$ limit
\begin{eqnarray}
\P({\bf 4_1};a,q=1,t)=a^{-2}t^{-2}+t^{-1}+1+t+at^2.
\end{eqnarray}
On the other hand, $x=1$ limit of the super-A-polynomial is listed in
table \ref{superAlimit}. Clearly the relation (\ref{superAsuperP}) holds
for the ${\bf 4_1}$ knot.

Let us now discuss properties of the figure-eight super-$A$-polynomial determined above.
First, we show that in appropriate limits it reproduces various known answers.
As expected, for $t=-1$ and $a=1$, the expression (\ref{Asuper41}) reduces to
\be
A(x, y) = (x-1)^2(y-1)\Big(x^2(y^2 + 1) - (1 - x - 2 x^2 - x^3 + x^4) y \Big),  \label{Afigure8}
\ee
which, apart from the $(x-1)^2$ factor, reproduces the $A$-polynomial of $K = {\bf 4_1}$,
including the $(y-1)$ factor representing the contribution of abelian flat connections.
We stress that both the factorization and the explicit form of this abelian branch is seen only in the limit $a=-t=1$
and is completely ``mixed'' with the other branches otherwise.
In general the super-$A$-polynomial (\ref{Asuper41}) does not factorize,
as is also the case for the unknot and torus knots that will be discussed next.

More generally, after a simple change of variables
\be
Q = a,\qquad \beta = x, \qquad \alpha = y\frac{1 - \beta Q}{Q(1-\beta)},
\label{super2av_fig8}
\ee
and for $t=-1$ we find that (\ref{Asuper41}) becomes
\bea
A^{\textrm{Q-def}}(\alpha,\beta,Q) & = & \frac{Q^2(1-\beta)^2}{\beta Q - 1}\, \Big( (\beta^2 - Q \beta^3) + (2 \beta - 2 Q^2 \beta^4 + Q^2 \beta^5 - 1) \alpha + \nonumber \\
& & + (1 - 2 Q \beta + 2 Q^2 \beta^4 - Q^3 \beta^5) \alpha^2 + Q^2 (\beta-1) \beta^2 \alpha^3 \Big) \,. \nonumber
\eea
Up to the first fraction, the expression in the big bracket reproduces the $Q$-deformed $A$-polynomial given in \cite{AVqdef}. A related change of variables, constructed along the lines of appendix \ref{ssec-Aaug}, reveals the form of closely related augmentation polynomial of \cite{NgFramed}.


\subsection{Trefoil knot}

In this section, we derive the classical and quantum super-$A$-polynomial for the trefoil knot (\emph{i.e.} $(2,3)$ torus knot, also denoted $T^{(2,3)}$ or ${\bf 3_1}$) and verify the validity of the Conjecture 1 and 2 for this knot.
The analysis follows the same lines as in previous sections, and its starting point is the expression for the colored superpolynomial.
We can provide such an expression from two sources. First, the colored superpolynomial
for general $(2,2p+1)$ torus knot was derived in \cite{FGS} from the perspective of the refined Chern-Simons theory.
This superpolynomial is given in~(\ref{Paqt-torus}), as we will need it for the analysis of general torus knots in the next section.
Even though in this section we only need $p=1$ specialization of (\ref{Paqt-torus}), this is still quite an intricate expression.
Nonetheless we are able to find its simpler form\footnote{Note that for $a=q^2$ and $t=-1$, the expression (\ref{Paqt31}) reduces to $\P_n({\bf 3_1};q^2,q,-1)=\sum_{k=0}^{n-1} q^{n(k+1)-1} (q^{n-1},q^{-1})_k $, which provides a simpler form of the Jones polynomial for the trefoil considered in \cite{Habiro,Garoufalidis}, $\P_n({\bf 3_1};q^2,q,-1)=J_n({\bf 3_1};q)=\sum_{k=0}^{n-1} (-1)^k q^{k(k+3)/2 +nk} (q^{-n-1},q)_k(q^{-n+1},q)_k $.}
\be
\P_n ({\bf 3_1};a,q,t) = \sum_{k=0}^{n-1}  a^{n-1} t^{2k} q^{n(k-1)+1} \frac{(q^{n-1},q^{-1})_k(-a t q^{-1},q)_k}{(q,q)_k} \,. \label{Paqt31}
\ee
Another way to derive this formula is to consider constraints arising from the action of various differentials,
as also discussed at length in \cite{FGS}. It turns out that these constraints also lead to the above formula
for the colored superpolynomial for the trefoil (even though they are much harder to analyze for other torus knots).
We also note that in \cite{FGS} a specialization $a=q^2$ of (\ref{Paqt31}) was used in order to find refined $A$-polynomial.
Here we use more general $a$-dependent expression with the goal of finding super-$A$-polynomial.
Explicit values of $\P_n ({\bf 3_1};a,q,t)$ following from (\ref{Paqt31}) for low values of $n$ are given in table~\ref{tab-P31}.

\begin{table}[h]
\centering
\begin{tabular}{|@{$\Bigm|$}c|@{$\Bigm|$}l|}
\hline
$\, n$ & $\qquad \P_n ({\bf 3_1};a,q,t)$  \\
\hline
\hline
$\, 1$ & $1$   \\
\hline
$\, 2$ & $ a q^{-1} + a q t^2  + a^2 t^3  $    \\
\hline
$\, 3$ & $ a^2 q^{-2} + a^2 q (1 + q) t^2 + a^3 (1 + q) t^3 + a^2 q^4 t^4 +
 a^3 q^3 (1 + q) t^5 + a^4 q^3 t^6$  \\
\hline
$\, 4$ & $  a^3 q^{-3} + a^3 q (1 + q + q^2) t^2 + a^4 (1 + q + q^2) t^3 +
 a^3 q^5 (1 + q + q^2) t^4 + $ \\
& $+ a^4 q^4 (1 + q) (1 + q + q^2) t^5 + a^3 q^4 (a^2 + a^2 q + a^2 q^2 + q^5) t^6 + $ \\
& $+  a^4 q^8 (1 + q + q^2) t^7 + a^5 q^8 (1 + q + q^2) t^8 + a^6 q^9 t^9$  \\
\hline
\end{tabular}
\caption{Colored superpolynomial of the ${\bf 3_1}$ knot for $n=1,2,3,4$.
\label{tab-P31} }
\end{table}

Again, from the explicit form of the colored superpolynomial (\ref{Paqt31}) we find the recursion
relation it satisfies by using the Mathematica package \textrm{qZeil.m}, see \cite{qZeil}. This recursion relation takes the form
\be
\hat A^{\text{super}} (\hat x, \hat y; a,q,t) = a_0 + a_1 \hat{y} + a_2 \hat{y}^2 \,,   \label{Ahat31}
\ee
where
\bea
a_0 & = & \frac{a^2 t^4 (\hat{x}-1) \hat{x}^3 (1 + a q t^3 \hat{x}^2)}{ q (1 + a t^3 \hat{x}) (1 + a t^3 q^{-1} \hat{x}^2) }   \nonumber \\
a_1 & = & -\frac{a (1 + a t^3 \hat{x}^2) \big( q - q^2 t^2 \hat{x} +t^2 (q^2 + q^3 + a t + a q^2 t) \hat{x}^2 + a q^2 t^5 \hat{x}^3 + a^2 q t^6 \hat{x}^4 \big)}{q^2 (1 + a t^3 \hat{x}) (1 + a t^3 q^{-1}\hat{x}^2) }  \nonumber \\
& & \qquad   \nonumber \\
a_2 & = &  1   \nonumber
\eea
At this point we also stress that the simple representation (\ref{Paqt31}) is essential for deriving this second order recursion relation;
the algorithm \textrm{qZeil.m} applied to the equivalent, but more involved expression (\ref{Paqt-torus}) finds only the sixth order recursion relation.

\begin{figure}[ht]
\begin{center}
\includegraphics[width=0.4\textwidth]{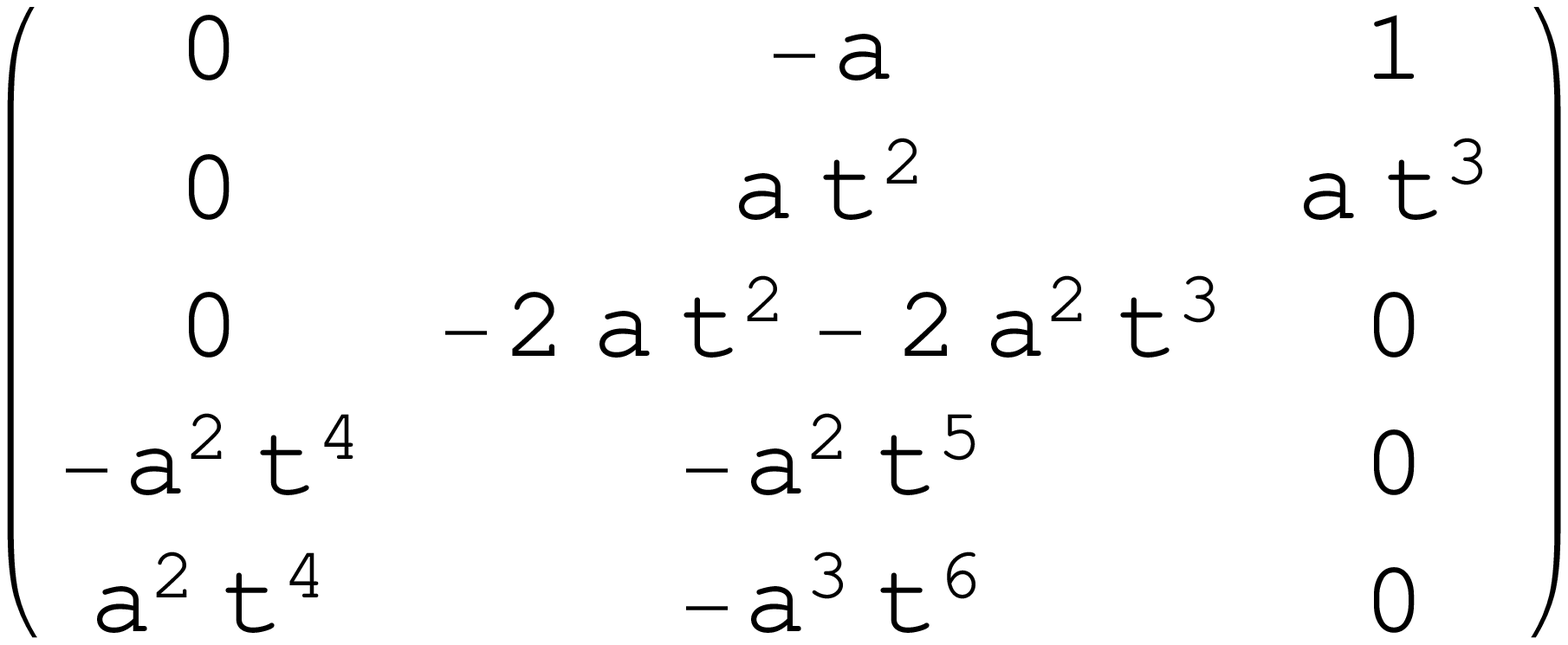}
\caption{Matrix form of the super-$A$-polynomial for the trefoil knot. The conventions are the same as in figure \protect\ref{fig-unknotNewtonMatrix}.}
\label{fig:matrix31}
\end{center}
\end{figure}

\begin{figure}[ht]
\begin{center}
\includegraphics[width=0.4\textwidth]{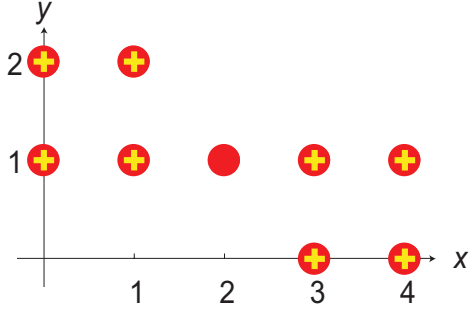}
\caption{Newton polygon of the super-$A$-polynomial for the trefoil knot and its $a=-t=1$ limit. The conventions are the same as in the unknot case in figure \protect\ref{fig-unknotNewtonMatrix}.}
\label{fig:Newton31}
\end{center}
\end{figure}

The classical super-$A$-polynomial for trefoil knot follows from the $q\to 1$ limit of $\hat A^{\text{super}}$ and reads
\bea
A^{\text{super}} (x, y; a,t) & = & a^2 t^4 (x-1) x^3 + (1 + a t^3 x) y^2 + \label{Asuper31} \\
& & -a\big( 1 - t^2 x + 2 t^2 (1 + a t) x^2 + a t^5 x^3 + a^2 t^6 x^4 \big) y  \,.  \nonumber
\eea
Matrix form of this polynomial is presented in figure \ref{fig:matrix31}, and its Newton polygon is shown in figure \ref{fig:Newton31}.

Let us now show that the same polynomial can be derived from the asymptotic behavior of the colored superpolynomial~(\ref{Paqt31}).
Using integral representation as in (\ref{Pn41-integral}),
\be
\P_n ({\bf 3_1};a,q,t) \; \sim \; \int dz\; e^{\frac{1}{\hbar}\left(\widetilde{\mathcal{W}}({\bf 3_1};z,x)+{\cal O}(\hbar)\right)},
\ee
with the potential
\bea
& & \widetilde{\mathcal{W}}({\bf 3_1};z,x)  =  - \frac{\pi^2}{6} +\big(\log z + \log a \big)\log x + 2(\log t) (\log z)   \label{V31}\\
& & \quad   + \Li_2( x z^{-1}) - \Li_2(x) + \Li_2(-a t) - \Li_2(-a t z) + \Li_2(z) \,,  \nonumber
\eea
and in the limit (\ref{reflimit}) with $z=e^{\hbar k}=const$, we find that equations (\ref{saddle_point41}) and (\ref{yV}) take the form
\be
\left\{\begin{array}{l} 1 = \frac{t^2 x (x - z_0) (1 + a t z_0)}{z_0 (z_0-1)}   \\
y =  \frac{a z_0^2 (x-1)}{(x - z_0)}   \end{array} \right.
\ee
Eliminating $z_0$ from these two equations we reproduce the super-$A$-polynomial (\ref{Asuper31}).
We conclude that both Conjecture~1 and Conjecture~2 hold true for the trefoil knot.

We also consider various limits of the super-$A$-polynomial (\ref{Asuper31}). For $t=-1$ and $a=1$ we get
\be
A (x, y) = -(x-1)(y-1)(y+x^3) \,,
\label{AAAtref}
\ee
which reproduces the well known $A$-polynomial for the trefoil, including the $y-1$ factor
associated with abelian flat connections (and the overall immaterial factor $x-1$).
More generally, under a change of variables (which in fact is $p=1$ specialization of more general identification (\ref{super2av_torus}) valid of $(2,2p+1)$ torus knots)
\be
Q=a,\quad \beta=x,\quad \alpha=yQ^{-1}\beta^{-6}\frac{1-Q\beta}{1-\beta} \, ,
\ee
and in $t=-1$ limit, (\ref{Asuper31}) reduces (up to an overall factor) to
\be
A(\alpha,\beta,Q) = (1 - Q \beta) + (\beta^3 - \beta^4 + 2 \beta^5 -2 Q \beta^5 - Q \beta^6 + Q^2 \beta^7) \alpha + (-\beta^9 + \beta^{10}) \alpha^2,
\ee
which perfectly reproduces the $Q$-deformed or augmentation polynomial for the trefoil knot found in \cite{NgFramed,AVqdef}. Relations between super-$A$-polynomial, $Q$-deformed $A$-polynomial and augmentation polynomial for torus knots are also discussed in much more detail in appendix \ref{ssec-Aaug}.

\subsection{$(2,2p+1)$ torus knots}

As the last class of examples we discuss the entire family of $(2,2p+1)$ torus knots, which are also denoted $T^{(2,2p+1)}$.
The $S^r$-colored superpolynomials for this family were determined from the refined Chern-Simons theory viewpoint in~\cite{FGS}.
{}From this perspective the reduced colored superpolynomial is found as the ratio of (refined) Chern-Simons partition functions
in $\S^3$ in presence of a given knot and the unknot.
(Recall, that the unreduced colored superpolynomial is given as a similar ratio of Chern-Simons partition functions
in the presence and absence of a knot, {\it cf.} (\ref{Punknot}).)
This computation, with all technicalities related to consistent refinement of Chern-Simons computation
(which involves taking into account appropriate $\gamma$-factors and other subtleties)
leads to the following expression for the colored superpolynomial for $(2,2p+1)$ torus knot:
\begin{eqnarray}
\P^{S^r} (T^{(2,2p+1)}; a,q,t)&=& \left( \frac{q_1}{q_2} \right)^{\tfrac{pr}{2}}
\frac{Z_{SU(N)}^{\text{ref}}(\S^3,T^{(2,2p+1)}_{\Lambda^r};q_1,q_2)}{Z_{SU(N)}^{\text{ref}}(\S^3,\unknot_{\Lambda^r};q_1,q_2)}  \label{Paqt-torus} \\
& = & \sum_{\ell=0}^r\frac{(qt^2;q)_{\ell}(-at^3;q)_{r+\ell}(-aq^{-1}t;q)_{r-\ell}(q;q)_{r}}
{(q;q)_{\ell}(q^2t^2;q)_{r+\ell}(q;q)_{r-\ell}(-at^3;q)_{r}}\frac{(1-q^{2\ell+1}t^2)}{(1-qt^2)}
\nonumber \\
&&\quad
\times (-1)^{n-1}a^{-\frac{r}{2}}q^{\frac{3(n-1)}{2}-\ell}t^{-(n-1)p-\ell+\frac{r}{2}}
\left[
(-1)^{\ell}a^{\frac{r}{2}}q^{\frac{r^2-\ell(\ell+1)}{2}}t^{\frac{3r}{2}-\ell}
\right]^{2p+1} \,.    \nonumber
\end{eqnarray}
Much as in the unknot example (\ref{Punknot}), at intermediate stages we need to compute partition functions
for anti-symmetric representations $\Lambda^r$, and the change of variables (\ref{GSvarchng}) has been performed
in order to present the final result in terms of $a,q,t$ variables; for more details see~\cite{FGS}.

We note that, for $p=1$, the expression (\ref{Paqt-torus}) can be rewritten in much simpler form (\ref{Paqt31}),
which leads to the desired second order recursion relation (\ref{Ahat31}),
while direct application of the algorithm \textrm{qZeil.m} to the representation (\ref{Paqt-torus})
with $p=1$ gives only sixth order relation.
This shows that recursion relations found directly from (\ref{Paqt-torus}) for arbitrary $p$ in general will not be of minimal order;
it is certainly interesting to find recursion relations of minimal order for all $p$.
Relegating this task to future work, in this section we focus on the asymptotic analysis
of (\ref{Paqt-torus}), which is sufficient for deriving the {\it classical} super-$A$-polynomials of minimal order for all $p$.
The knowledge of these classical super-$A$-polynomials is an important hint in deriving quantum super-$A$-polynomials of minimal order for general $p$.

In the asymptotic limit $\hbar\to 0$ limit, the colored superpolynomial $\P^{S^r}(T^{(2,2p+1)}; a,q,t)$ behaves as
\begin{eqnarray}
\P^{S^r} (T^{(2,2p+1)}; a,q,t)
\; \sim \; \int dz\; e^{\frac{1}{\hbar}\left(\widetilde{\mathcal{W}} (T^{(2,2p+1)};z,x)+{\cal O}(\hbar)\right)},
\end{eqnarray}
with the potential
\begin{eqnarray}
&& \widetilde{\mathcal{W}} (T^{(2,2p+1)};z,x) =
 p \log (a) \cdot \log x
-p\log (-t)\cdot \log x+(p+1)\pi i\log x+\log (x^{\frac{1}{2}}z^{-1})\cdot\log t
\nonumber \\
&&\quad\quad\quad\quad\quad\quad\quad
+(2p+1)\Biggl(
\pi i\log z+\frac{1}{2}\left((\log x)^2-(\log z)^2\right)
+\log (x^{\frac{3}{2}}z^{-1})\cdot \log t\Biggr)
\nonumber \\
&&\quad\quad\quad\quad\quad\quad\quad
+{\rm Li}_2(z)-{\rm Li}_2(x)-{\rm Li}_2(t^2z)+{\rm Li}_2(-at^3x)
+{\rm Li}_2(t^2xz)
\nonumber \\
&&\quad\quad\quad\quad\quad\quad\quad
-{\rm Li}_2(-at^3xz)
+{\rm Li}_2(xz^{-1})
-{\rm Li}_2(-atxz^{-1})
+{\rm Li}_2(-at)-{\rm Li}_2(1),
\end{eqnarray}
where $z = q^{\ell}$.

For the above potential $\widetilde{\mathcal{W}} (T^{(2,2p+1)};z,x)$,
the critical point condition can simply be expressed as $1=\exp\left(z\partial \widetilde{\mathcal{W}}/\partial z\right)|_{z=z_0}$:
\begin{eqnarray}
1 \; = \; -\frac{t^{-2-2p}(x-z_0)z_0^{-1-2p}(-1+t^2z_0)(1+ at^3 xz_0)}{(-1+z_0)(atx+z_0)(-1 +  t^2 x z_0)} \,.
\label{braid_saddle1}
\end{eqnarray}
and
\begin{eqnarray}
y(x,t,a)&=&\exp\left(x\frac{\partial \widetilde{\mathcal{W}} (T^{(2,2p+1)};z_0,x)}{\partial x}\right)
\nonumber \\
&=&\frac{a^p t^{2 + 2 p} (-1 + x) x^{1 + 2 p} (atx + z_0) (1 + a t^3 x z_0)}{(1 + a t^3 x) (x - z_0) (-1 + t^2 x z_0)} \,.
\label{braid_saddle2}
\end{eqnarray}
Eliminating $z_0$ from the above equations, we find the
super-$A$-polynomial $A^{\text{super}} (x,y;a,t)$ for any $(2,2p+1)$ torus knot.
For small values of $p$, the resulting super-$A$-polynomials are listed
in tables \ref{table_super},  \ref{table_super4} and
\ref{table_super5}.\footnote{In these tables, we omitted the extra
factors which appears in the elimination, and picked up the factor that
includes the non-abelian branch of the $SL(2)$ character variety. }
\begin{table}[h]
\centering
\begin{tabular}{|@{$\Bigm|$}c|@{$\Bigm|$}l|}
\hline
\textbf{Knot}  & $A_K^{\text{super}} (x,y;a,t)$  \\
\hline
\hline
$T^{(2,3)}$ & {\scriptsize $y^2 +$}$\frac{1}{1+ a t^3 x}${\scriptsize
     $a(-1 + t^2 x -2 t^2 x^2 - 2 a t^3 x^2 - a t^5 x^3 - a^2 t^6 x^4)y
 +$}$\frac{(x-1)a^2 t^4 x^3}{1 + a t^3 x}$    \\
\hline
$T^{(2,5)}$ & {\scriptsize $y^3
-$}$\frac{a^2}{1 + a t^3 x}$ {\scriptsize $
(1 - t^2 x + 2 t^2 x^2 + 2 a t^3 x^2 - 2 t^4 x^3 - 2 a t^5 x^3 +
 3 t^4 x^4 + 4 a t^5 x^4 + a^2 t^6 x^4 + a t^7 x^5
$
}
\\
&{\scriptsize
$\quad\quad
- a^2 t^8 x^5 +
 2 a^2 t^8 x^6)y^2$
}
\\
&{\scriptsize $+$}$\frac{a^4t^6 (-1 + x) x^5}{(1 + a t^3 x)^2}$
{\scriptsize $(
2 - t^2 x + a t^3 x + 3 t^2 x^2 + 4 a t^3 x^2 + a^2 t^4 x^2 +
 2 a t^5 x^3 + 2 a^2 t^6 x^3 + 2 a^2 t^6 x^4
$
}
\\
&{\scriptsize
$\quad
+ 2 a^3 t^7 x^4 +
 a^3 t^9 x^5 + a^4 t^{10} x^6
)y$
}
\\
&{\scriptsize $-$}$\frac{a^6t^{12} (-1 + x)^2 x^{10}}{(1 + a t^3 x)^2}$  \\
\hline
$T^{(2,7)}$ &
{\scriptsize $y^4
-$}$
\frac{a^3}{1+at^3x}$
{\scriptsize
$(1 - t^2 x + 2 t^2 x^2 + 2 a t^3 x^2 - 2 t^4 x^3 - 2 a t^5 x^3 +
  3 t^4 x^4 + 4 a t^5 x^4 + a^2 t^6 x^4 - 3 t^6 x^5$}
\\&{\scriptsize $\quad\quad
- 4 a t^7 x^5 -
  a^2 t^8 x^5 + 4 t^6 x^6 + 6 a t^7 x^6 + 2 a^2 t^8 x^6 + a t^9 x^7 -
  2 a^2 t^{10} x^7 + 3 a^2 t^{10} x^8
)y^3
$}
\\&{\scriptsize $+$}$\frac{a^6t^8 (-1 + x) x^7}{(1 +a t^3 x)^2}$
{\scriptsize $(
3 - 2 t^2 x + a t^3 x + 6 t^2 x^2 + 8 a t^3 x^2 + 2 a^2 t^4 x^2 -
 3 t^4 x^3 - 2 a t^5 x^3 + a^2 t^6 x^3 + 6 t^4 x^4
$}\\
&{\scriptsize $\quad
+ 12 a t^5 x^4 +
 10 a^2 t^6 x^4 + 4 a^3 t^7 x^4 + 3 a t^7 x^5 + 2 a^2 t^8 x^5 -
 a^3 t^9 x^5 + 6 a^2 t^8 x^6 + 8 a^3 t^9 x^6 + 2 a^4 t^{10} x^6
$}\\
&{\scriptsize $\quad
+  2 a^3 t^{11} x^7 - a^4 t^{12} x^7 + 3 a^4 t^{12} x^8)y^2
$}
\\
&{\scriptsize $-$}$\frac{a^9t^{16} (-1 + x)^2 x^{14}}{(1 +a t^3 x)^3}$
{\scriptsize
$
(3 - t^2 x + 2 a t^3 x + 4 t^2 x^2 + 6 a t^3 x^2 + 2 a^2 t^4 x^2 +
  3 a t^5 x^3 + 4 a^2 t^6 x^3 + a^3 t^7 x^3
$}\\
&{\scriptsize $\quad
+ 3 a^2 t^6 x^4 +
  4 a^3 t^7 x^4 + a^4 t^8 x^4 + 2 a^3 t^9 x^5 + 2 a^4 t^{10} x^5 +
  2 a^4 t^{10} x^6 + 2 a^5 t^{11} x^6 + a^5 t^{13} x^7 + a^6 t^{14} x^8)y$
}
\\
&{+$\frac{a^{12}t^{24} (-1 + x)^3 x^{21}}{(1 + t^3 x)^3}$}  \\
\hline
\end{tabular}
\caption{
Super-$A$-polynomials for $(2,2p+1)$ torus knots with $p=1,2,3$. For  cases $p=4,5$ see tables
 \protect\ref{table_super4} and \protect\ref{table_super5}.\label{table_super} }
\end{table}

For $a=1$ we find the refined $A$-polynomials of \cite{FGS}. On the other hand,
for $t=-1$ we find the $Q$-deformed $A$-polynomial of \cite{AVqdef}
if the following identification of parameters is performed
\begin{eqnarray}
Q=a,\quad \beta=x,\quad \alpha=yQ^{-p}\beta^{-4p-2}\frac{1-Q\beta}{1-\beta} \, ,
\label{super2av_torus}
\end{eqnarray}
and a related transformation reveals the form of the augmentation polynomial of \cite{NgFramed}.
Precise derivation of the above variable change, as well as explicit relations between super-$A$-polynomial, the augmentation polynomial and $Q$-deformed $A$-polynomial, are discussed in detail in appendix \ref{ssec-Aaug}.

For $n=2$ the colored superpolynomial (\ref{Paqt-torus}) becomes
\begin{eqnarray}
\P(T^{(2,2p+1)};a,q,t)
=\frac{a^pq^{-p}(1+aqt^3-q^{2p+1}t^{2p+2}(q+at))}{1-q^2t^2}.
\end{eqnarray}
For $q=1$, the superpolynomial reduces to
\begin{eqnarray}
\P(T^{(2,2p+1)};a,q=1,t)
=\frac{a^p(1+at^3-t^{2p+2}(1+at))}{1-t^2}.
\label{super-A_torus_red}
\end{eqnarray}
Up to $p=5$, we can compare this limit and the super-A-polynomials
reduced at $x=1$ which are given in table \ref{superAlimit}, and  
the relation (\ref{superAsuperP}) can be confirmed explicitly.


\section{Quantizability}
\label{sec:quantizability}

In this section we discuss the super-$A$-polynomials that we found from the viewpoint of quantizability,
by which we mean the following. For the Conjecture~1 to be formulated in a consistent way,
we must ensure that the leading term $\int \log y \frac{dx}{x}$ in the integral (\ref{VCsuper}) is well-defined,
{\it i.e.} does not depend on the choice of the integration path on the algebraic curve~(\ref{supercurve}).
As explained in~\cite{Apol,abmodel},
this requirement imposes the following constraints on the periods of the imaginary and real parts of $\log y \frac{dx}{x}$, respectively,
\bea
\oint_{\gamma} \Big( \log |x| d ({\rm arg} \, y) - \log |y| d ({\rm arg} \, x) \Big) & = & 0 \,, \label{qcond0} \\
\frac{1}{4 \pi^2} \oint_{\gamma} \Big( \log |x| d \log |y| + ({\rm arg} \, y) d ({\rm arg} \, x) \Big) & \in & \mathbb{Q} \, ,
\label{qcondQ}
\eea
for all {\it closed} paths $\gamma$ on the curve (\ref{supercurve}).
It turns out that these conditions can be further reformulated and interpreted in a variety of ways.
On one hand, it is amusing to observe that the integrand $\eta(x,y)=\log |x| d ({\rm arg} \, y) - \log |y| d ({\rm arg} \, x)$
in \eqref{qcond0} is the image of the symbol $\{ x,y \} \in K_2 (\cC)$ under so-called regulator map,
thereby constituting an immediate link to algebraic K-theory  \cite{Beilinson,Bloch,Fernando}.
As discussed in~\cite{abmodel}, from this K-theory viewpoint the condition that the curve is quantizable
can be rephrased simply as the requirement that $\{ x,y \} \in K_2 (\C (\cC))$ is a torsion class.
On the other hand, this more abstract condition also translates to the down-to-earth statement
that quantizability of the curve requires its defining polynomial to be {\it tempered}.

By definition, a polynomial $A(x,y)$ is tempered if all roots of all face polynomials of its Newton polygon are roots of unity.
Face polynomials are constructed as follows: we need to construct a Newton polygon
corresponding to $A(x, y) = \sum_{i,j} a_{i,j}x^i y^j$, and to each point $(i, j)$ of this polygon we associate the coefficient $a_{i,j}$.
We label consecutive points along each face of the polygon by integers $k=1,2,\ldots$ and,
for a given face, rename monomial coefficients associated to these points as $a_k$.
Then, the face polynomial associated to a given face is defined to be $f(z)=\sum_k a_k z^k$.
Therefore, the quantizability condition requires that all roots of $f(z)$ constructed
for all faces of the Newton polygon must be roots of unity. In what follows we are going
to examine super-$A$-polynomials which we found in examples in section \ref{sec-case} from this perspective.

Ordinary $A$-polynomials have numerical, integer coefficients \cite{CCGL}, and therefore the above
quantization condition imposes certain constraints on values of these coefficients.
For example, the ordinary $A$-polynomial of the figure-eight knot given in (\ref{Afigure8}) satisfies these constraints,
while its close cousin with only slightly different coefficients, discussed {\it e.g.} in \cite{Apol,abmodel}, does not.
Meeting these tight constraints might seem much less trivial in the case of $t$- or $a$-deformed curve,
when coefficients of the defining polynomial depend on these extra parameters.
Nonetheless, we found in \cite{FGS} that this is indeed possible for refined ({\it i.e.} $t$-dependent) $A$-polynomials
and the outcome is very simple: the quantization condition implies that $t$ has to be a root of unity.
Therefore, even though such $t$ can not be completely arbitrary, it still takes values in a dense set of points (on a unit circle).
With this result in mind, the reader should not be surprised that an analogous conclusion applies
to the super-$A$-polynomial as well: all super-$A$-polynomials found in section~\ref{sec-case}
are tempered (and therefore quantizable) as long as both $a$ and $t$ are roots of unity.
Moreover, this condition very nicely fits with the fact that in specialization from colored superpolynomial
or HOMFLY polynomial to $sl(N)$ quantum group invariant we substitute $a=q^N$
and in Chern-Simons theory with $SU(N)$ gauge group $q$ is required to be a root of unity,
so that $a=q^N$ is automatically a root of unity as well!

\begin{table}[ht]
\centering
\begin{tabular}{| c || c | }
\hline
\rule{0pt}{5mm}
face & face polynomial \\[3pt]
\hline
\rule{0pt}{5mm}
N  & $ z + a t $ \\[3pt]
NE & $ z + a t^2 $  \\[3pt]
E  & $ a t^2(z+a t^3)^2$    \\[3pt]
SE & $ a^3 t^8(z - a t^2)$  \\[3pt]
S  & $ a^3 t^9 (z + a t) $  \\[3pt]
SW & $ a^2 t^5(z-a t^4) $  \\[3pt]
W  & $ a^2 t^5(z - 1)^2 $  \\[3pt]
NW & $ a t (z-a t^4) $  \\[3pt]
\hline
\end{tabular}
\caption{Face polynomials for the figure-eight knot, corresponding to faces of the octagonal shape formed by non-zero entries of the coefficient matrix in figure \protect\ref{fig:matrix41}. Faces are labeled by compass directions (with N standing for North, \emph{etc}.), with the first row $(0,-at,-1,0)$ of the matrix in figure \protect\ref{fig:matrix41} located in the North.}   \label{tab-face-figure8}
\end{table}

Let us now illustrate the above claim in the examples of various knots discussed in section~\ref{sec-case}.
For each of those knots we construct a Newton polygon and face polynomials of the corresponding super-$A$-polynomials.
In order to construct face polynomials it is convenient to write down a matrix representation of the super-$A$-polynomials.
For instance, for the unknot the Newton polygon and the corresponding matrix representation
are shown in figure~\ref{fig-unknotNewtonMatrix}. In this case, it is clear that roots of face polynomials
are all roots of unity if $a$ and $t$ are roots of unity.
In fact, the unknot is so simple that even a weaker condition is sufficient to hold, namely that the combination $at^3$ is a root of unity.

The matrix coefficients and the Newton polygon for figure-eight knot are given, respectively,
in figures \ref{fig:matrix41} and \ref{fig:Newton41}, and the corresponding face polynomials are presented in table~\ref{tab-face-figure8}.
The face polynomials are manifestly written as products of linear factors, and being tempered requires that both $a$ and $t$ are roots of unity.

An analogous condition holds for $(2,2p+1)$ torus knots.
In particular, the matrix coefficients and the Newton polygon for the trefoil knot are shown,
respectively, in figures \ref{fig:matrix31} and~\ref{fig:Newton31}.
Newton polygons of other $(2,2p+1)$ torus knots have a similar, hexagonal shape which grows linearly with $p$;
in fact, they are identical to Newton polygons for the refined $A$-polynomials presented in~\cite{FGS}.
The face polynomials of the super-$A$-polynomials discussed in the present paper are certain
$a$-deformations of the refined face polynomials studied in~\cite{FGS},
and they also factorize into linear factors as shown in table~\ref{tab-face-torus}.
By inspection, it is clear that roots of these polynomials are all roots of unity as long as both $a$ and $t$ are roots of unity.

To sum up, we conclude that super-$A$-polynomials which we find are quantizable if both $a$ and $t$ are roots of unity
and we conjecture that this is the case for all knots. Let us note that this condition is also consistent with
what we found in~\cite{FGS} in the context of refined mirror curves for BPS states.
For example, in that paper we discussed in detail the refined mirror curve for the conifold,
which apart from $t$ variable depends also on the K{\"a}hler parameter $Q$.
By a similar analysis as here we found that both $t$ and $Q$ have to be roots of unity.
On the other hand, in the relation between string theory and Chern-Simons theory the K{\"a}hlar parameter $Q$
is interpreted as the variable $a$ of knot polynomials, and from various other viewpoints
({\it e.g.} quantum foam, crystal models, attractor mechanism, {\it etc.}) the K{\"a}hlar parameter
is also quantized as $Q=q^N$ with $N$ being the rank of the Chern-Simons gauge group.
Therefore, we see that both knot theory and BPS state perspectives lead to similar and consistent conclusions.

\begin{table}[ht]
\centering
\begin{tabular}{| c || c | }
\hline
\rule{0pt}{5mm}
face & face polynomial \\[3pt]
\hline
\rule{0pt}{5mm}
first column & $-(a t^2)^{p(p+1)} (z-1)^p$ \\[3pt]
last column & $(-1)^p (z+a t^3)^p$  \\[3pt]
first row & $z a^p -1$    \\[3pt]
last row & $-(a t^2)^{p(p+1)} \big(z-(a t^2)^{p} \big)$  \\[3pt]
lower diagonal & $(-1)^p  \big(a t^3)^{p} (z - a^{p+1}t^{2p+1} \big)^p$  \\[3pt]
upper diagonal & $(-1)^{p+1} a^p \big(z + a^p t^{2p+2} \big)^p$  \\[3pt]
\hline
\end{tabular}
\caption{Face polynomials for $(2,2p+1)$ torus knots, corresponding to faces of the hexagonal shape formed by non-zero entries of the coefficient matrices for $(2,2p+1)$ torus knots, such as the matrix for the trefoil in figure \protect\ref{fig:matrix41}.}   \label{tab-face-torus}
\end{table}


\section{Differentials}
\label{sec:diff}

We wish to emphasize one important point which did not affect our examples
and, therefore, was suppressed in our discussion so far.
It has to do with various specializations of the super-$A$-polynomial illustrated
in figure~\ref{fig:superAlimits} or, to be more precise, with analogous specializations
of its quantum version that encode recursion relations for various $S^r$-colored knot invariants, see table~\ref{AAAtable}.

For example, both $t=-1$ specializations in figure~\ref{fig:superAlimits} have a clear ``quantum'' analog,
which corresponds to passing from homological to polynomial knot invariants.
In the triply-graded case, this operation of taking graded Euler characteristic relates
(the Poincar\'e polynomial of) the colored HOMFLY homology to the colored HOMFLY polynomial,
whereas in the doubly-graded cases it relates (the Poincar\'e polynomial of) the $n$-colored
Khovanov homology
to the colored Jones polynomial $J_n (K;q)$, see {\it e.g.} figure~2 of~\cite{FGS}.

In particular, because the Poincar\'e polynomials of these homology theories are related
to the corresponding knot polynomials by the specialization $t=-1$, the same is true
about recursion relations that describe the ``color behavior'' of these invariants:
\bea
\hat A^{\text{super}} (\hat x, \hat y; a,q,t) \vert_{t = -1} & = & \hat A^{\text{Q-def}} (\hat x, \hat y; a,q) \\
\hat A^{\text{ref}} (\hat x, \hat y; q,t) \vert_{t = -1}  & = & \hat A (\hat x, \hat y; q). \nonumber
\eea
Similarly, a specialization $\hat A^{\text{Q-def}} (\hat x, \hat y; a,q) \vert_{a=q^2} = \hat A (\hat x, \hat y; q)$
presents no difficulty since evaluation of the $n$-colored  HOMFLY polynomial at $a=q^2$ gives
the $n$-colored Jones polynomial for all values of $n$.
Therefore, any recursive relation on the former is guaranteed to yield a recursion relation
for $J_n (K;q)$ via specialization to $a=q^2$.

On the other hand, the specialization \eqref{AsuperAref} is more delicate since, in general,
the colored superpolynomial $\P_n (K; a,q,t)$ evaluated at $a=q^2$
is {\it not} equal to the Poincar\'e polynomial of the colored Khovanov homology, $\P^{sl(2),V_n} (q,t)$,
where $V_n = S^{n-1}$ is the $n$-dimensional representation of $sl(2)$.
Instead, the two homology theories are related by a certain differential~\cite{GS},
which at the level of Poincar\'e polynomials implies a relation\footnote{In the grading conventions
of \cite{DGR,GWalcher,AS,DMMSS}, the factor on the right-hand side reads $(1 + a^{-2} q^{4} t^{-1})$.}
\be
\P_n (a,q,t) \; = \;
R_n (a,q,t) + (1 + a^{-1} q^{2} t^{-1}) Q_n (a,q,t) \,,
\label{superright}
\ee
where $R_n (a,q,t)$ and $Q_n (a,q,t)$ are polynomials with non-negative coefficients,
such that $\P^{sl(2),V_n} (q,t) = R_n (q^2,q,t)$.
Hence, when the differential acts non-trivially, a (recursion) relation among
colored superpolynomials $\P_n (a,q,t)$ does {\it not} automatically lead to a recursion relation
for $sl(2)$ Poincar\'e polynomials $\P^{sl(2),V_n} (q,t)$ due to the extra terms $Q_n (a,q,t) \ne 0$.
In other words, unless $\hat A^{\text{super}} (\hat x, \hat y; a,q,t)$
annihilates $(1 + a^{-1} q^{2} t^{-1}) Q_* (a,q,t)$ at $a=q^2$,
it will {\it not} produce an operator $\hat A^{\text{ref}} (\hat x, \hat y; q,t)$
by a simple rule \eqref{AsuperAref}:
\be
\hat A^{\text{super}} (\hat x, \hat y; q^2,q,t) \, \P^{sl(2),V_*} (q,t) \; + \;
(1 + t^{-1})  \hat A^{\text{super}} (\hat x, \hat y; q^2,q,t) \, Q_* (q^2,q,t) \; = \; 0 \,.
\ee
Note, this issue does not exist in the unrefined (``decategorified'') case since setting $t=-1$ makes the second term vanish.

This is the only subtle specialization in the ``quantum'' version of the diagram in figure~\ref{fig:superAlimits}.
We expect, however, that even this subtlety goes away in the classical limit $q \to 1$.
Indeed, all commutative deformations of the $A$-polynomial describe ``large color''
behavior of various knot polynomials, {\it cf.} \eqref{VCsuper}.
In particular, we expect that the number of generators in the $n$-colored HOMFLY homology
killed by the differential \eqref{superright} exhibits slower than exponential growth in the large-$n$ limit \eqref{reflimit},
\be
\lim_{n \to \infty} \, \frac{1}{n} \log \, Q_n (a,q,t) \; = \; 0 \,,
\ee
and, therefore, does not muddy the waters in the diagram
in figure~\ref{fig:superAlimits}.\\

Besides playing a key role in various specializations of homological knot invariants,
the differentials endow knot homologies with a very rich structure,
which turns out to be very elegant and often so constraining that one can even compute
colored superpolynomials based on this structure alone, with a minimal input.
In particular, this is how nice formulas like \eqref{Paqt31} can be produced.
Referring the reader to \cite{GS} for further details, here we merely state
a simple rule of thumb: the factors of the form $(1 + a^i q^j t^k)$ that we often
see {\it e.g.} in
\eqref{Punrec},
\eqref{Paqt41},
\eqref{Paqt31}, and \eqref{Paqt-torus}
come from differentials of $(a,q,t)$-degree $(i,j,k)$, {\it cf.}~\cite[eq. (3.54)]{FGS}:
\be
\begin{array}{c@{\;}|@{\;}c@{\;}|@{\;}c@{\;}c}
\text{differentials} & \text{factors} & (a, q, t)~\text{grading} \\\hline
d_{N>0} & \quad 1 + a q^{-N} t \quad & (-1,N,-1) \\[.1cm]
d_{N<0} & 1 + a q^{-N} t^3 & (-1,N,-3) \\[.1cm]
d_{\text{colored}} & 1 + q & (0,1,0) \\[.1cm]
& 1 + at & (-1,0,-1) \\
& \vdots &
\end{array}
\label{gradingtabl}
\ee
For example, notice that all terms with $k>0$ in the expression \eqref{Paqt41}
for the colored superpolynomial of the figure-eight knot
manifestly contain a factor $(1 + a q^{n-1} t^3)$.
Hence, the $S^{n-1}$-colored superpolynomial of the figure-eight knot
has the following structure, {\it cf.} \eqref{superright}:
\be
\P_n ({\bf 4_1};a,q,t) \; = \; 1 + (1 + a q^{n-1} t^3) Q_n (a,q,t) \,,
\ee
which means that, when evaluated at $a = - q^{1-n} t^{-3}$,
the sum \eqref{Paqt41} collapses to a single $k=0$ term,
$\P_n ({\bf 4_1}; a = - q^{1-n} t^{-3},q,t) = 1$.
A proper interpretation of this fact is that a specialization to $N = 1-n$
of the triply-graded $S^{n-1}$-colored HOMFLY homology, carried out by the action of the differential $d_{1-n}$, is trivial.
In other words, the differential $d_N$ with $N=1-n$ is canceling in a theory with $R = S^{n-1}$.

The reason for this is very simple: specialization to $N<0$ is best understood --- in view
of the ``mirror symmetry'' \cite{GS} --- as a specialization to $sl(-N)$ knot homology
colored by a transposed Young diagram $R^t$.
In the present case, it means that the $\Lambda^{n-1}$-colored $sl(N)$ knot homology is
trivial ({\it i.e.} one-dimensional for every knot $K$) when $N=n-1$, which, of course,
must be the case since the representation $R^t = \Lambda^{N}$ of $sl(N)$ is trivial.
Similarly, one can explain much of the structure that {\it a priori} may seem random
in the colored superpolynomials and the super-$A$-polynomials, or even derive them.


\section{Physical interpretation}
\label{sec:phys}

In our previous work, we proposed to interpret the parameter $t$ responsible for
the ``refinement'' or ``categorification'' as a twisted mass parameter for the global symmetry $U(1)_F$
in the effective three-dimensional $\cN=2$ theory $T_M$ associated to the knot complement $M = \S^3 \setminus K$:
\be
M \quad \leadsto \quad T_M \,.
\label{TMfromM}
\ee
Moreover, generically, every charged chiral multiplet in a theory $T_M$ contributes to
the effective twisted chiral superpotential a dilogarithm term:
\be
\text{chiral field}~\phi
\qquad \longleftrightarrow \qquad
\begin{array}{l}
\text{twisted superpotential} \\[.1cm]
\Delta \widetilde{\cal W} (\vec x; t) = \textrm{Li}_2
\Big( (-t)^{n_F} \prod_i (x_i)^{n_i} \Big)
\end{array}
\label{xtsuper}
\ee
where $n_F$ is the charge of the chiral multiplet under the global R-symmetry $U(1)_F$
and $\{ n_i \}$ is our (temporary) collective notation for all other charges of $\phi$ under symmetries $U(1)_i$,
some of which may be global flavor symmetries and some of which may be dynamical gauge symmetries,
depending on the problem at hand.\footnote{Below we shall return to the different role of gauge and global symmetries,
but for now we wish to point out a simple rule of thumb that one can read off the matter content
of the theory $T_M$ by counting dilogarithm terms in the function $\widetilde{\cal W} (\vec x; t)$.}
In particular, in the former case, the vev of the corresponding twisted chiral multiplet
is usually called the twisted mass parameter $\tilde m_i = \log x_i$,
of which $\tilde m_F = \log (-t)$ is a prominent example.

\begin{figure}[ht]
\centering
\includegraphics[width=2.6in]{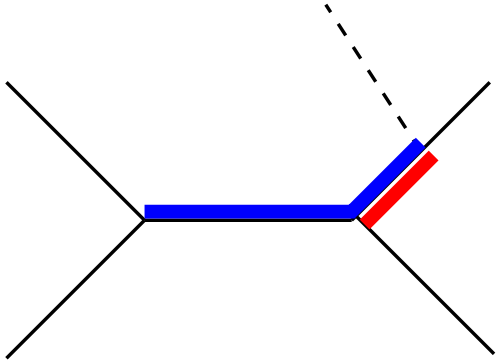}
\caption{A toric Lagrangian brane in the conifold bounds two holomorphic disks
(shown by red and blue intervals in the base of the toric geometry).}
\label{fig:rbbrane}
\end{figure}

The second commutative deformation parameter $a$ also admits a similar interpretation
as a twisted mass parameter for a global symmetry that we denote $U(1)_Q$:
\be
\log a \; = \; \tilde m_Q \,.
\label{mftrel}
\ee
In fact, in the case of the $a$-deformation this interpretation is even more obvious
and can be easily seen in the brane picture, where it corresponds to one of the K\"ahler moduli
of the underlying Calabi-Yau geometry $X$.
For example, the effective low-energy theory on a toric brane in the conifold geometry
has two chiral multiplets that come from two open BPS states shown in blue and red in figure~\ref{fig:rbbrane}.
We stress that this brane picture can be directly seen from our results for the unknot: the $a$-deformed curve
(\ref{Aunknot1x1y}) represents a genuine conifold mirror curve $A=1-ax-\tilde{y}+x\tilde{y}$, after a simple rescaling $\tilde{y}=a^{1/2}y$ and with K{\"a}hler parameter $a$.
In this case the structure of two dilogarithms arises as the leading order term in the asymptotic expansion of $t=-1$ specialization of the superpolynomial (\ref{Punknot}) (which is essentially given by the ratio of quantum dilogarithms, as is the case for the brane amplitude in the conifold geometry).

\bigskip

In this example, the symmetry $U(1)_Q$ responsible for the $a$-deformation comes from
the 2-cycle in the conifold geometry $X$. (The corresponding gauge field $A_{\mu}$
comes from the Kaluza-Klein reduction of the RR 3-form field, $C \sim A \wedge \omega$,
and becomes the starting point for the geometric engineering of $\cN=2$ gauge theories in four dimensions \cite{engineering}.)
In a basis of refined open BPS states shown in figure~\ref{fig:rbbrane},
one state is charged under the symmetry $U(1)_Q$, while the other state is neutral.
Therefore, the effective twisted superpotential $\widetilde{\cal W} (x; a, t)$ of the corresponding
model has two dilogarithm terms, one of which depends on $a$ and the other does not.

Returning to the general theory $T_M$, now we are ready to explain the connection between
the twisted superpotential in this theory and the algebraic curve \eqref{supercurve}
defined as the zero locus of the super-$A$-polynomial.
Roughly speaking, the curve \eqref{supercurve} describes the SUSY vacua in the $\cN=2$ theory $T_M$.
To make this more precise, we need to recall that among the parameters $x_i$ in \eqref{xtsuper}
some correspond to vevs of dynamical fields (and, therefore, need to be integrated out)
and some are twisted masses for global flavor symmetries.
To make the distinction clearer, let us denote the former by $z_i$ (instead of $x_i$),
so that the vevs of dynamical twisted chiral superfields are $\sigma_i = \log z_i$.
Then, in order to find SUSY vacua of the theory $T_M$ we need to extremize $\widetilde{\cal W}$
with respect to these dynamical fields,
\be
\frac{\partial \widetilde{\cal W}}{\partial z_i} \; = \; 0 \,.
\label{wzextr}
\ee
This is exactly what we did {\it e.g.} in \eqref{saddle_point41} when we extremized the potential function \eqref{V41}
for the figure-eight knot ({\it cf.} also \eqref{V31} and \eqref{braid_saddle1} for the case of $(2,2p+1)$ torus knots).
Solving these equations for $z_i$ and substituting the resulting values back into $\widetilde{\cal W}$
gives the effective twisted superpotential, $\widetilde{\cal W}_{\text{eff}}$, that depends only on twisted mass
parameters associated with global symmetries of the $\cN=2$ theory $T_M$.

Besides the symmetries $U(1)_F$ and $U(1)_Q$ which are responsible for $t$- and $a$-deformations, respectively,
our $\cN=2$ theories $T_M$ come with additional global flavor symmetries, one for each component of the link $K$
(or, more generally, one for every torus boundary of $M$).
In particular, if $K$ is a knot --- which is what we assume throughout the present paper --- then,
in addition to $U(1)_F$ and $U(1)_Q$, there is only one extra global symmetry $U(1)_L$
with the corresponding twisted mass parameter that we simply denote $\tilde m$;
it is $x = e^{\tilde m}$ that shortly will be identified with the variable by the same name in the super-$A$-polynomial.
In the brane model,
\be
\begin{matrix}
{\mbox{\rm space-time:}} & \qquad & \R^4 & \times & X \\
& \qquad & \cup &  & \cup \\
{\mbox{\rm D4-brane:}} & \qquad & \R^2 & \times & L
\end{matrix}
\label{surfeng}
\ee
this symmetry $U(1)_L$ can be identified with the gauge symmetry on the D4-brane supported on the Lagrangian
submanifold $L \subset X$. The corresponding gauge field is dynamical when $L$ has finite volume,
while for non-compact $L$ (of infinite volume) the symmetry $U(1)_L$ is a global symmetry.
Moreover, the other global symmetry $U(1)_F$ that plays an important role
in our discussion also can be identified in the brane setup \eqref{surfeng}:
it corresponds to the rotation symmetry of the normal bundle of $\R^2 \subset \R^4$.

To summarize our discussion so far, we can incorporate $U(1)_Q$ and $U(1)_L$ charges in \eqref{xtsuper}
and write the contribution of a chiral multiplet $\phi \in T_M$ to the twisted superpotential as
\be
\text{chiral field}~\phi
\qquad \longleftrightarrow \qquad
\begin{array}{l}
\text{twisted superpotential} \\[.1cm]
\Delta \widetilde{\cal W} (x, z_i; a, t) = \textrm{Li}_2
\Big( a^{n_Q} (-t)^{n_F} x^{n_L} \prod_i (z_i)^{n_i} \Big)
\end{array}
\label{axtsuper}
\ee
Using this dictionary and dilogarithm identities, such as the inversion formula
$\textrm{Li}_2 (x) = - \textrm{Li}_2 \left( \frac{1}{x} \right) - \frac{\pi^2}{6} - \frac{1}{2} [\log (-x)]^2$, from \eqref{V41}
and \eqref{V31} it is easy to read off
the spectrum of the theory $T_M$ for the trefoil knot and for the figure-eight knot:

\begin{table}[htb]
\be
\begin{array}{l@{\;}|@{\;}ccccc@{\;}|@{\;}c}
\multicolumn{7}{c}{\text{trefoil knot}} \\[.1cm]
& \phi_1 & \phi_2 & \phi_3 & \phi_4 & \phi_5 & \text{parameter} \\\hline
U(1)_{\text{gauge}} & -1 & 0 & 0 & -1 & 1 & z \\
U(1)_{F}            & 0 & 0 & 1 & -1 & 0 & -t \\
U(1)_{Q}            & 0 & 0 & 1 & -1 & 0 & a \\
U(1)_{L}            & 1 & -1 & 0 & 0 & 0 & x
\end{array}
\qquad\qquad
\begin{array}{l@{\;}|@{\;}ccccccc}
\multicolumn{8}{c}{\text{figure-eight knot}} \\[.1cm]
& \phi_1 & \phi_2 & \phi_3 & \phi_4 & \phi_5 & \phi_6 & \phi_7 \\\hline
U(1)_{\text{gauge}} & 0 & -1 & 0 & -1 & 0 & -1 & -1 \\
U(1)_{F}            & 0 & 0 & 1 & -1 & 3 & -3 & 0 \\
U(1)_{Q}            & 0 & 0 & 1 & -1 & 1 & -1 & 0 \\
U(1)_{L}            & -1 & 1 & 0 & 0 & 1 & -1 & 0
\end{array}
\notag \ee
\caption{Spectrum of the $\cN=2$ theory $T_M$ for the trefoil and figure-eight knots.}
\label{tab:charges}
\end{table}

\noindent
The terms of lower transcendentality degree, {\it i.e.} products of ordinary logarithms,
also admit a simple interpretation in three-dimensional $\cN=2$ gauge theory $T_M$.
Notice that, in the collective notations $\{ x_i \}$ for global and gauge symmetries $U(1)_i$
used in \eqref{xtsuper}, the dependence of the twisted superpotential $\widetilde{\cal W}$
on $\log x_i$ is always quadratic, see {\it e.g.} \eqref{V41} and \eqref{V31}.
Such terms correspond to supersymmetric Chern-Simons couplings for $U(1)$ gauge (resp. background flavor) fields:
\be
\frac{k_{ij}}{4 \pi} \int A_i \wedge d A_j + \ldots
\qquad \longleftrightarrow \qquad
\begin{array}{l}
\text{twisted superpotential} \\[.1cm]
\Delta \widetilde{\cal W} (\vec x; a,t) = \frac{k_{ij}}{2} \, \log x_i \cdot \log x_j
\end{array}
\label{superCSTM}
\ee
At this point, we should remind the reader that a given $\cN=2$ theory $T_M$ may admit many dual UV descriptions,
with different number of gauge groups and charged matter fields \cite{DGG}.
However, all of these dual descriptions lead to the same space of supersymmetric moduli (twisted mass parameters)
once all dynamical multiplets are integrated out,
{\it i.e.} once the twisted superpotential is extremized \eqref{wzextr} with respect to all $z_i$.

The resulting ``effective'' twisted superpotential $\widetilde{\cal W}_{\text{eff}} (x;a,t)$ depends
only on the twisted mass parameters associated with the global symmetries $U(1)_L$, $U(1)_Q$, and $U(1)_F$.
Then, the algebraic curve \eqref{supercurve} defined as the zero locus of the super-$A$-polynomial
is simply a graph of the function $x \frac{\partial \widetilde{\cal W}_{\text{eff}}}{\partial x}$,
which in a circle compactification of the theory $T_M$ is interpreted as the effective FI parameter:
\be
\boxed{\phantom{\oint} \cM_{\text{SUSY}}: \quad A^{\text{super}} (x,y;a,t) = 0
\qquad \Leftrightarrow \qquad  \log y = x \partial_{x} \widetilde{\cal W}_{\text{eff}} (x; a, t)
\phantom{\oint} }
\label{aviaw}
\ee
See \cite{DGH,DGSdual} for a similar interpretation of the ordinary $A$-polynomial.
The only difference is that, in the present discussion, we also keep track of the $U(1)_F$ and $U(1)_Q$ quantum numbers,
which result in the $a$- and $t$-dependence of the twisted superpotential.



\section{Concluding remarks}
\label{sec:rems}

It is important to realize a few key points that make this story work and allow us to define a 2-parameter
family deformation of the classical $A$-polynomial.

The first point, realized already in \cite{FGS}, is that the parameter $t$ responsible for categorification
in knot theory applications is a {\it commutative} deformation parameter. In other words, unlike \eqref{xycomm},
it does not change the algebra of functions in $x$ and $y$, which under the $t$-deformation remains commutative.
To appreciate the importance of this point, consider a generating function, $Z_{\text{ref. BPS}}^{\text{open}} (X,L)$,
of refined open BPS invariants for a Lagrangian brane in a (toric) Calabi-Yau $X$.
To characterize this function --- which can be fairly involved --- it is often convenient to focus on
a difference equation that it obeys, rather than on a function itself.
Writing this difference equation in the form of a Schr\"odinger-like equation {\it a la}~\eqref{AonZ},
\be
\hat A \; Z_{\text{ref. BPS}}^{\text{open}} (X,L) \; = \; 0 \,,
\ee
and shifting the focus to the operator $\hat A$ is often a smart way to encode the information
contained in $Z_{\text{ref. BPS}}^{\text{open}} (X,L)$.
In the unrefined case ({\it i.e.} when $q_1 = q_2$), the fact that brane partition functions obey such
Schr\"odinger-like equations goes back to the pioneering work \cite{ADKMV} on
integrable hierarchies and topological strings (see also \cite{DHSV,DHS,ACDKV}), whose fully refined version,
with generic values of both parameters $q_1$ and $q_2$ was studied only recently in \cite{FGS}.

\begin{figure}[ht]
\centering
\includegraphics[width=3.6in]{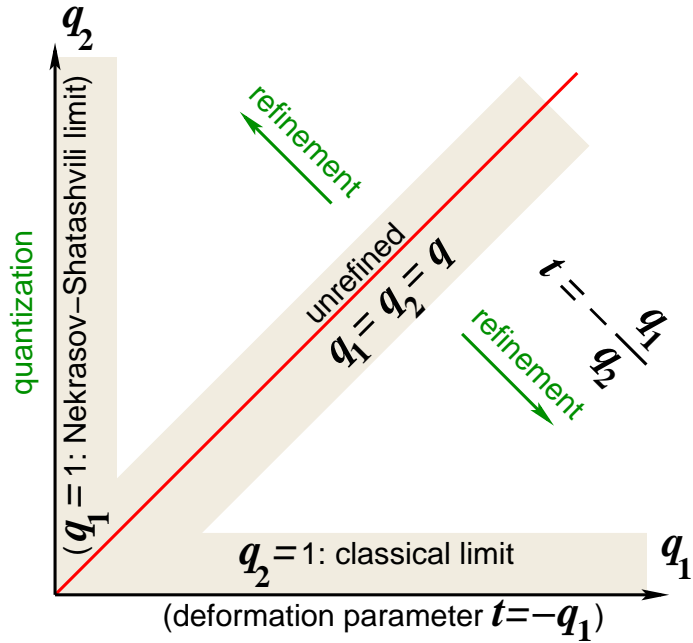}
\caption{With our choice of conventions, the parameter $q_2 = q = e^{\hbar}$ is responsible for
quantization, whereas $t = - \frac{q_1}{q_2}$ is the deformation parameter responsible for the refinement.
Hence, the ``classical limit'' corresponds to $q_2=1$, while $q_1=1$ is the so-called Nekrasov-Shatashvili limit.
Finally, $q_1 = q_2$ defines a locus in the space of parameters where refinement is turned off,
and the problem can be formulated in terms of ordinary topological strings.}
\label{fig:qqqt}
\end{figure}

Even if on general grounds one believes that {\it refined} brane amplitudes should be annihilated by operators $\hat A (\hat x, \hat y)$,
there is no {\it a priori} reason why the refinement parameter $t = - \frac{q_1}{q_2}$ should not appear in
the algebra of $\hat x$ and $\hat y$. Indeed, it is clear that only a certain combination of parameters $q_1$ and $q_2$,
say $f(q_1,q_2)$, can enter the commutation relation
\be
\hat y \hat x \; = \; f(q_1,q_2) \, \hat x \hat y \,,
\label{xyfcomm}
\ee
but {\it a priori} $f(q_1,q_2)$ could be a non-trivial function of both $q_1$ and $q_2$.
However, a closer look at the physics of refined BPS states quickly shows that $f(q_1,q_2)$ is equal to either $q_1$ or $q_2$,
depending on how one orients the brane in space-time.
With our choice of conventions, it is the parameter $q_2$ that enters the commutation relation \eqref{xyfcomm} in the refined context,
and luckily the relation to knot theory variables $q$ and $t$ turns out to be just right \cite{FGS}:
\be
q \; = \; q_2 \,, \qquad
t \; = \; - \frac{q_1}{q_2} \,.
\label{qqqt}
\ee
so that when \eqref{xyfcomm} is expressed in terms of $q$ and $t$, the parameter $t$ does {\it not} appear
in the commutation relation of $\hat x$ and $\hat y$, and therefore plays the role of the {\it commutative} deformation parameter.
This explains why the ``classical'' limit \eqref{reflimit} of refined / homological invariants is simply $q \to 1$
with fixed $t$, as opposed to a more general limit of the form
\be
q^p t^q \to 1 \,, \qquad q^r t^s = \text{fixed} \,,
\ee
with some ``exponents'' $p$, $q$, $r$ and $s$. Note, with this choice of conventions,
the so-called Nekrasov-Shatashvili limit \cite{NS} corresponds to studying {\it closed} refined BPS invariants
and setting $q_1=1$, see figure \ref{fig:qqqt}.
Expansion of the {\it open} refined brane amplitude around this limit was recently studied in \cite{ACDKV},
where it was argued that, at least in some class of examples, the ``quantum'' curve $\hat A (\hat x, \hat y, q_2)$ has the same
form as the classical curve $A(x,y)$, {\it i.e.} is independent of the parameter $q_2$ as long as $q_1 = 1$, see also \cite{Robbert}.
It would be interesting to study this further and, in particular, to see if the full quantum super-$A$-polynomial $\hat A^{\text{super}}$
happens to coincide with the classical super-$A$-polynomial $A^{\text{super}}$ in this class of examples.

\bigskip

The second important point, recently emphasized in \cite{AVqdef}, provides a similar explanation for the second
commutative deformation parameter $a$ which, when combined with the $t$-deformation,
gives us the desired 2-parameter family studied in this paper.
Indeed, since the highest weight of a $SU(N)$ representation $R$ has $N-1$ components,
{\it a priori} one might expect that ``color behavior'' of $sl(N)$ polynomial or homological
knot invariants, such as $J_R (K;q)$ or $\P^{sl(N), R} (K;q,t)$, is captured by
a variety\footnote{{\it cf.} \cite[sec. 2.3]{DGreview} for a higher-rank version of the generalized volume conjecture} of dimension $N-1$.
In particular, when $N=2$ this variety is one-dimensional, defined by a single polynomial $A(x,y)$,
which has to do with the fact that all irreducible representations of $SU(2)$ are labeled by a single integer $n = \dim R$.

On the other hand, when $N>2$ the weight space has dimension $N-1$ and the classical
limit of recursion relations for $J_R (K;q)$ or $\P^{sl(N), R} (K;q,t)$ is a higher-dimensional algebraic variety,
generically defined by $N-1$ equations,
\be
\cM_K : \quad A_i (\vec x, \vec y) \; = \; 0
\qquad i = 1, \ldots, N-1\,,
\label{assocVar}
\ee
in the ``phase space'' $\left( \C^* \times \C^* \right)^{N-1} / S_N$ of total dimension $2N-2$.
The standard arguments from Chern-Simons theory show that this variety is Lagrangian with
respect to the holomorphic symplectic form $\Omega_J = \frac{i}{\hbar} \sum_{i=1}^{N-1} \frac{dx_i}{x_i} \wedge \frac{dy_i}{y_i}$,
thereby endowing it with the structure of the so-called $(A,B,A)$ brane, in the terminology of \cite{Kapustin-Witten}.
This curious fact has many far-reaching applications and consequences \cite{surf-op-braids},
but our interest here is merely in a simple fact that one of the defining polynomials in \eqref{assocVar}
is essentially the $A$-polynomial of the $sl(2)$ theory.
Indeed, if we restrict our attention to totally symmetric representations,
then the system \eqref{assocVar} collapses to a single equation, $A^{\text{Q-def}}(x,y;a)=0$,
which must contain the ordinary $A$-polynomial as a factor since symmetric representations is all one has for $N=2$.
Furthermore, it was argued in \cite{AVqdef} that the $Q$-deformed $A$-polynomial $A^{\text{Q-def}}(x,y;a)$
is equal to the augmentation polynomial of knot contact homology \cite{NgFramed}.


Finally, let us make a few remarks on the physical interpretation of the super-$A$-polynomial $A^{\text{super}} (x,y;a,t)$.
Even though the interpretation as a space of SUSY vacua described in section \ref{sec:phys} opens many doors for further study,
it would still be interesting to understand the interpretation of $a$- and $t$-deformations in a good old Chern-Simons theory.
Such interpretation is not of a purely academic interest: it could help to understand better the role of
framing dependence in knot homologies which, up to present day, has been very mysterious.
Understanding how framing affects homological knot invariants can help to reconcile different approaches
to colored knot homologies, see \cite[sec. 6.2]{GS} for some discussion on this point.



\acknowledgments{We thank M.~Aganagic, R.~Dijkgraaf, E.~Gorsky,
A.~Mironov, A.~Morozov, L.~Ng,
M.~Sto$\check{\text{s}}$i$\acute{\text{c}}$, and C.~Vafa for useful discussions on related topics.
The work of H.F. is supported by the Grant-in-Aid for Young Scientists
(B) [\# 21740179] from the Japan Ministry of Education, Culture, Sports,
Science and Technology, and the Grant-in-Aid for Nagoya University
Global COE Program, ``Quest for Fundamental Principles in the Universe:
from Particles to the Solar System and the Cosmos.''
The work of S.G. is supported in part by DOE Grant DE-FG03-92-ER40701FG-02 and in part by NSF Grant PHY-0757647.
The research of P.S. is supported by the DOE grant DE-FG03-92-ER40701FG-02,
the European Commission under the Marie-Curie International Outgoing Fellowship Programme, and the Foundation for Polish Science.
Opinions and conclusions expressed here are those of the authors and do not necessarily reflect the views of funding agencies.}


\appendix

\section{Colored superpolynomial for figure-eight knot}   \label{app-figure8}

In this appendix we present how the formula for $S^r$-colored superpolynomial for figure-eight knot (\ref{Paqt41}) arises (with $n=r+1$), and discuss how its specialization relates to the colored HOMFLY polynomial found in \cite{kawagoe}. The formula (\ref{Paqt41}) arises essentially from rewriting of the expression conjectured in \cite{ItoyamaMMM}, which is postulated as follows.\footnote{In typing this appendix we did our best to avoid several misprints which are present in the formulas in \cite{ItoyamaMMM}.} Firstly, as observed in \cite{DMMSS} in some examples and conjectured to be true for every knot $K$, the so-called ``special'' polynomial, i.e.  $q_{I}\to 1$ limit of the HOMFLY polynomial $P^R(K;A_{I},q_{I})$ colored by representation $R$, has the following property
\be
\lim_{q_{I}\to 1 }P^R(K;A_{I},q_{I}) = \Big( \lim_{q_{I}\to 1 } P^{[1]}(K;A_{I},q_{I}) \Big)^{|R|}.  \label{PR-1}
\ee
Note that here we use variables $A_{I}$ and $q_{I}$ which are to be identified with $A$ and $q$ used in \cite{ItoyamaMMM}. In what follows we also use the notation
$$
\{ x \} = x - x^{-1}, \qquad  [x] = \frac{q_I^x - q_I^{-x}}{q_I - q_I^{-1}}.
$$
In case of the figure-eight knot it is known that $P^{[1]}({\bf 4_1};A_{I},q_{I}) = 1 + \{ A \}^2$, so that (\ref{PR-1}) takes form
\be
\lim_{q_{I}\to 1 }P^R({\bf 4_1};A_{I},q_{I}) = \Big( 1 + \{ A \}^2 \Big)^{|R|} = \sum_{k=0}^{|R|} \frac{|R|!}{k! (|R|-k)!}  \{ A \}^{2k}.
\ee
It was postulated in \cite{ItoyamaMMM} that to introduce the full $q_I$-dependence of the (reduced) colored HOMFLY polynomial one should simply replace ordinary numbers in the above expression by their $q$-deformations. In particular, for symmetric representation $S^r$ represented by a Young diagram $[r]$ consisting of one row of $r$ boxes, this deformation takes the following form
\be
P^{[r]}({\bf 4_1};A_{I},q_{I})  = \sum_{k=0}^{r} \frac{[r]!}{[k]! [r-k]!} \prod_{i=0}^{k-1} \{ A_I q_I^{r+i} \} \{A_I q_I^{i-1} \}.
\ee
This expression we would like further deform by introducing $t_I$ dependence, which would encode Poincar{\'e} polynomials of homological figure-eight knot invariants. To this end we rewrite the above expression in the form
\be
P^{[r]}({\bf 4_1};A_{I},q_{I})  = \sum_{k=0}^{r} \sum_{1\leq i_1 < i_2 < \ldots < i_k \leq r} Z_{i_1}(A_I) Z_{i_2}(A_i q_I) Z_{i_3}(A_I q_I^2) \cdots Z_{i_k}(A_I q_I^{k-1}),  \label{Pr41Z}
\ee
where
$$
Z_i(A_I) = \{ A_I q_I^{2(r-i)+1} \} \{A_I q_I^{-1} \}.
$$
The expression (\ref{Pr41Z}) looks like a summation over boxes in a Young diagram, with contributions from each box being given by a function of its arm-length or leg-length. The proposal of \cite{ItoyamaMMM} is to introduce a familiar generalization of this type of formula, by representing leg-length and arm-length contributions respectively by $q_I$- and $t_I$-dependent expressions, which can be achieved by a substitution
\be
Z_i(A_I) \rightarrow \zeta_i(A_I)  =  \{ A_I q_I^{2(r-i)+1} \} \{A_I t_I^{-1} \} = Z_i(A_I)\frac{\{A_I t_I^{-1} \}}{\{A_I q_I^{-1} \}}.
\ee
Now we notice that (\ref{Pr41Z}), with $Z_{i}$ replaced by $\zeta_i$, can be rewritten as
\bea
P^{[r]}({\bf 4_1};A_{I},q_{I},t_I)  &=& \sum_{k=0}^{r} \sum_{1\leq i_1 < i_2 < \ldots < i_j \leq r} \zeta_{i_1}(A_I) \zeta_{i_2}(A_i q_I) \zeta_{i_3}(A_I q_I^2) \cdots \zeta_{i_k}(A_I q_I^{k-1})  \nonumber \\
& = & \sum_{k=0}^{r} \Big(\prod_{s=0}^{k-1} \frac{\{A_I q_I^s t_I^{-1} \}}{\{A_I q_I^{s-1} \}}  \Big) \sum_{1\leq i_1 < i_2 < \ldots < i_k \leq r} Z_{i_1}(A_I) Z_{i_2}(A_i q_I) \cdots Z_{i_k}(A_I q_I^{k-1})  \nonumber \\
& = &  \sum_{k=0}^{r} \Big(\prod_{s=0}^{k-1} \frac{\{A_I q_I^s t_I^{-1} \}}{\{A_I q_I^{s-1} \}}  \Big) \frac{[r]!}{[k]! [r-k]!} \prod_{i=0}^{k-1} \{ A_I q_I^{r+i} \} \{A_I q_I^{i-1} \}  \nonumber \\
& = & \sum_{k=0}^{r} \frac{[r]!}{[k]! [r-k]!} \prod_{i=0}^{k-1} \{A_I q_I^i t_I^{-1} \} \{ A_I q_I^{r+i} \}  \nonumber \\
& = & \sum_{k=0}^{r} \prod_{j=0}^{k-1}\Big(  -t_I A_I^{-2} q_I^{1-2j}\; \frac{1 - q_I^{2j - 2r}}{1-q_I^{2j+2}} (1 - A_I^2 q_I^{2j}t_I^{-2}) (1 - A_I^2 q_I^{2r+2j}) \Big).  \nonumber
\eea
This is the expression which we have been after. Now a change of variables from $A_I,q_I,t_I$ to $a,q,t$, given in (\ref{AIqItI-aqt}), leads to the final formula (\ref{Paqt41}) (where we use $n=r+1$, and the range of summation can be trivially extended to infinity).

\bigskip

For completeness let us also state the formula for the unreduced colored HOMFLY polynomial given by K. Kawagoe in \cite{kawagoe} and used in \cite{AVqdef}. This formula is written in grading conventions consistent with our notation, so that in terms of our variables $a$ and $q$ it takes form
\bea
\P^{\textrm{Kawagoe}}_n({\bf 4_1};a,q) & = & \frac{(a,q)_{n-1}}{(q,q)_{n-1}} \sum_{i=0}^n \sum_{j=0}^i  (-1)^i a^{\frac{2i+3-3n}{2}}  q^{\frac{2j+i^2-5i-5 + (4i+7)n - 2n^2 }{2}} \times \label{PnKawagoe} \\
& & \times \frac{\big( (q^{-i},q)_j (q^{1-n},q)_i \big)^2 (a q^{n-i+j-1},q)_{i-j} }{(q,q)_i (q,q)_j (q^{n-i+j},q)_{i-j}}  \nonumber
\eea
We verified that this formula agrees with $t=-1$ specialization of our expression $\P_n({\bf 4_1};a,q,t)$ given in (\ref{Paqt41}), up to the unknot normalization $\bar{\P}_n (\unknot; a,q,t)$ given in (\ref{Punknot}) and up to the overall sign
\be
\P^{\textrm{Kawagoe}}_n({\bf 4_1};a,q) = (-1)^{n-1} \, \bar{\P}_n (\unknot; a,q,-1) \,  \P_n({\bf 4_1};a,q,-1).  \label{Pour-Kawagoe}
\ee


\section{Knot contact homology and augmentation polynomial} \label{KCH}

The knot contact homology describes knot invariants as invariants of
the Legendrian submanifolds in the contact manifold
\cite{Chekanov, EES, EFM, ENS, Ng_comp, Sabloff, Ng1, Ng2, NgSabloff, NgFramed, Ng, EENS_rev}.
In particular, the geometric set-up of the knot contact homology in
\cite{Ng_top_string, EENS}
is similar to the topological A-model on $T^*M$
\cite{OoguriV,Koshkin,Taubes}, and the string-theoretical interpretation
of the knot contact homology has been recently studied in \cite{AVqdef}.
The contact structure can be introduced by replacing the cotangent
bundle $T^*M$ by the cosphere bundle $ST^*M$,
so that the knot is realized by an intersection with the unit conormal
bundle $L_K$ \cite{Ng_top_string, EENS}.
To extract the Legendrian isotopy invariant of $L_K$, the framework of
Legendrian contact homology \cite{Eliashberg} and
the Symplectic Field Theory \cite{SFT} can be applied, and knot
invariants such as the $A$-polynomial can be obtained by the transverse knot
contact homology \cite{NgFramed,Ng}.

In \cite{Ng1,Ng2,NgFramed, Ng}, the invariants of the knot contact homology $HC_*(K)$ have been
studied in terms of the differential graded algebra ${\mathcal A}$
(abbreviated as ``DGA''), which was introduced first in \cite{Chekanov}.
The DGA is a pair $({\mathcal A},\partial)$ which consists of the tensor
algebra ${\cal A}$ with grading and differential $\partial$ lowering the
degree by 1 unit, such that $\partial^2=0$. The differential is defined in terms of the automorphism $\phi\in {\text Aut}({\mathcal A})$.

In a combinatorial formulation of the knot contact homology \cite{Ng1},
the braid group $B_n$ defines the automorphism $\phi$ for the knot DGA
$Aut({\mathcal A}_n)$ for the tensor algebra ${\cal A}_n$ which consists
of $n(n-1)$ generators $a_{ij}$ ($1\le i,j\le n; i\ne j$).
Accompanying the higher degree generators,  the DGA is freely generated by
\begin{eqnarray}
&&\{a_{ij}\}_{1\le i,j\le n; i\ne j} \quad {\rm degree}\;0,
\\
&&\{b_{ij}\}_{1\le i,j\le n; i\ne j} \quad {\rm degree}\;1,
\nonumber \\
&&\{c_{ij}\}_{1\le i,j\le n} \quad {\rm degree}\;1,
\nonumber \\
&&\{e_{ij}\}_{1\le i,j\le n} \quad {\rm degree}\;2.
\nonumber
\end{eqnarray}
We denote ${\bf A}$ as the matrix $(a_{ij})$,
where we set $a_{ii} = -2$ for all $i$, and similarly for ${\bf B}:=(b_{ij})$,
${\bf C}:=(c_{ij})$, ${\bf D}:=(d_{ij})$.
For example, the $(2,2p+1)$ torus knot is described by $n=2$ and braid group element $B=\sigma_1^{2p+1}$.

In the above basis, the automorphism action $\sigma_k$ ($k=1,\cdots, n-1$)
for the braid group generator $\sigma_k\in B_n$ reads \cite{Ng1}
\begin{eqnarray}
\phi_{\sigma_k}:\left\{
\begin{array}{ccll}
a_{ki} & \mapsto & -a_{k+1,i} - a_{k+1,k}a_{ki} & i\ne k,\; k + 1 \\
a_{ik} & \mapsto & -a_{i,k+1} - a_{ik}a_{k,k+1} & i\ne k,\; k + 1 \\
a_{k+1,i} & \mapsto & a_{ki} & i\ne k,\; k + 1 \\
a_{i,k+1} & \mapsto & a_{ik} & i\ne k,\; k + 1 \\
a_{k,k+1} & \mapsto & a_{k+1,k} &  \\
a_{k+1,k} & \mapsto & a_{k,k+1} &  \\
a_{ij} & \mapsto & a_{ij} & i,j\ne k,\; k + 1
\end{array}
\right.
\end{eqnarray}
In the definition of the differential,
the extension map $\phi^{\text{ext}}:B_n\hookrightarrow B_{n+1}\to
{\text Aut}({\cal A}_{n+1})$ is used which is
the faithful representation of $B_n$. The inclusion $B_n\hookrightarrow B_{n+1}$
is obtained by adding the $(n + 1)^{th}$ strand,
which does not interact with other $n$ strands.
The automorphism $\phi^{\text{ext}}_B$ acts on $a_{i,n+1}$ and $a_{n+1,i}$
($i=1,\cdots n$) as $n\times n$ matrices $\Phi_B^{L}$ and $\Phi_B^{R}$
\begin{eqnarray}
\phi_B^{\text{ext}}(a_{i,n+1})=\sum_{\ell=1}^n(\Phi_B^{L})_{i\ell}a_{\ell,n+1},
\quad
\phi_B^{\text{ext}}(a_{n+1,i})=\sum_{\ell=1}^na_{n+1,\ell}(\Phi_B^{R})_{\ell i}.
\end{eqnarray}

The differential $\partial$ on ${\mathcal A}_n$ is defined by
\begin{eqnarray}
\partial {\bf A} &=& 0,
\nonumber \\
\partial {\bf B} &=& (1-\Phi_B^L)\cdot {\bf A},
\nonumber \\
\partial {\bf C} &=& {\bf A}\cdot (1-\Phi_B^R)
\\
\partial {\bf D} &=& {\bf B}\cdot (1-\Phi_B^R)-(1-\Phi_B^L)\cdot {\bf
 C},
\nonumber \\
\partial e_i &=& ({\bf B}+\Phi_B^L\cdot {\bf C})_{ii}.
\nonumber
\end{eqnarray}
In this way, the knot DGA is defined by a pair $({\mathcal
A}_n,\partial)$, and
the degree-0 transverse homology $HT_0(B)$ is given by
\begin{eqnarray}
HT_0(B)={\cal A}_n/\bigl(
(1-\Phi_B^L)\cdot {\bf A},\;{\bf A}\cdot (1-\Phi_B^R)
\bigr).
\end{eqnarray}

The knot DGA can also be generalized to the algebra over the ring
$R=\mathbb{Z}[\lambda^{\pm},\mu^{\pm}]$, see \cite{NgFramed}.
Furthermore, in \cite{Ng} the tensor algebra
over $R[U,V]$ has been introduced. In this more general framework
$\phi_B$ and $\phi^{\text{ext}}$ are not changed, however ${\bf A}$ is
generalized to a pair consisting of $\hat{\bf A}$ and $\check{\bf A}$, such that
\begin{eqnarray}
(\hat{\bf A})_{ij}=\left\{
\begin{array}{cc}
a_{ij} & i>j \\
-1-\mu U & i=j \\
\mu U a_{ij} & i<j
\end{array}
\right.
\quad
(\check{\bf A})_{ij}=\left\{
\begin{array}{cc}
Va_{ij} & i>j \\
-V-\mu  & i=j \\
\mu  a_{ij} & i<j
\end{array}
\right.
\end{eqnarray}
The degree-0 infinity transverse homology of $B$
is given by
\begin{eqnarray}
HT_0^{\infty}(B)=({\mathcal A}_n\otimes R[U^{\pm 1},V^{\pm 1}])/
(\hat{\bf A}-{\bf \Lambda}_B'\cdot \Phi_B^L\cdot \check{\bf A},\;
\check{\bf A}-\cdot \hat{\bf A}\cdot\Phi_B^R\cdot ({\bf \Lambda}_B')^{-1}
),
\label{homology0}
\end{eqnarray}
where
\begin{eqnarray}
{\bf \Lambda}_B'={\rm diag}(\lambda\mu^{-w(B)}(U/V)^{-({\rm sl}(B)+1)/2},1,\cdots,1),
\end{eqnarray}
and $w(B)$ and ${\rm sl}(B)$ denote respectively the writhe and self-linking number.

To extract the topological information of the knot $K$ out of
$HT^{\infty}$, we can consider so-called augmentation polynomial.
The augmentation polynomial for $HT^{\infty}(K)$ is defined as the
resultant of the ideal ${\mathcal I}_K$ which comes from
$\hat{\bf A}-{\bf \Lambda}_B'\cdot \Phi_B^L\cdot \check{\bf A}=0$, and
$\check{\bf A}-\cdot \hat{\bf A}\cdot\Phi_B^R\cdot ({\bf \Lambda}_B')^{-1}=0$.
Here we assume the abelian basis $a_{ij}$ for tensor algebra ${\cal A}_n$.
To reduce the number of variables, we use one of the conditions of the full DGA, which reads
\begin{eqnarray}
{\bf A}_0-{\bf \Lambda}_B'\cdot\phi_K({\bf A}_0)\cdot({\bf \Lambda}_B')^{-1}=0,
\label{DGA_B}
\end{eqnarray}
where $({\bf A}_0)_{ij}=a_{ij}$ and $({\bf A}_0)_{ii}=0$.
Eliminating extra variables and taking the resultant, we find the augmentation polynomial
$\text{Aug}_K(\mu,\lambda;U,V)$. As we will discuss in what follows, this polynomial is closely related to and generalizes the $A$-polynomial.

\subsection{Augmentation polynomial for $(2,2p+1)$ torus knots}  \label{ssec-aug}

As an example, in this section we study the knot contact homology and
derive corresponding augmentation polynomials for $(2,2p+1)$ torus knots, see table \ref{table_aug}.
While these results follow (implicitly) from the analysis in \cite{Ng2,NgFramed, Ng},
we find it useful to write the resulting augmentation polynomials explicitly,
and relate them directly to $t=-1$ specialization of super-$A$-polynomials for torus knots
which we listed in tables \ref{table_super}, \ref{table_super4} and
\ref{table_super5}.\footnote{The Mathematica package for computing the transverse homology is available from Lenhard Ng's website \cite{transverse}.}

We recall that $(2,2p+1)$ torus knots can be constructed from the $2p+1$ braidings between 2 strands.
Therefore the DGA for $B_2$ gives the knot contact homology for this class of
torus knots. For $n=2$ the action of $\phi$ yields simply \cite{Ng1}
\begin{eqnarray}
\phi_{\sigma_1}:\;a_{12}\mapsto a_{21},\quad  a_{21}\mapsto a_{12},
\end{eqnarray}
and $\phi^{\text{ext}}_{\sigma_1}$ acts as
\begin{eqnarray}
\phi^{\text{ext}}_{\sigma_1}:
\left\{
\begin{array}{l}
a_{1*}\mapsto -a_{2*}-a_{21}a_{1*},\quad
a_{2*}\mapsto a_{1*},
\\
a_{*1}\mapsto -a_{*2}-a_{1*}a_{12},\quad
a_{*2}\mapsto a_{*1},
\end{array}
\right.
\end{eqnarray}
where we denoted $*:=n+1=3$.
The action of $\phi^{\text{ext}}_{\sigma_1^{2p+1}}$ is obtained by iterating the
above action $2p+1$ times. The matrices for $\Phi^{L,R}_{\sigma_1^{2p+1}}$
($p=1,2,3$) are found as:\\
$\bullet$ $p=1$:
\begin{eqnarray}
&&
\Phi_{\sigma_1^3}^L=\left(
\begin{array}{cc}
2 a_{21} - a_{21} a_{12} a_{21} & 1-a_{21}a_{12} \\
-1+a_{12}a_{21} & a_{12}
\end{array}
\right),\;
\Phi_{\sigma_1^3}^R=\left(
\begin{array}{cc}
2 a_{12} - a_{12} a_{21} a_{12} & -1+a_{12}a_{21} \\
-1+a_{21}a_{12} & a_{21}
\end{array}
\right).
\end{eqnarray}
$\bullet$ $p=2$:
\begin{eqnarray}
&&
\Phi_{\sigma_1^5}^L=\left(
\begin{array}{cc}
-3 a_{21} +4 a_{21} a_{12} a_{21} - a_{21} a_{12} a_{21} a_{12} a_{21}
& -1+3a_{21}a_{12} - a_{21} a_{12} a_{21} a_{12}  \\
1-3a_{12}a_{21} -a_{12} a_{21} a_{12} a_{21}  & -2a_{12}+a_{12}a_{21}a_{12}
\end{array}
\right),
\nonumber \\
&&
\Phi_{\sigma_1^5}^R=\left(
\begin{array}{cc}
-3 a_{12} +4 a_{12} a_{21} a_{12}- a_{12} a_{21} a_{12} a_{21} a_{12}
& 1-3a_{12}a_{21} + a_{12} a_{21} a_{12} a_{21} \\
-1+3a_{21}a_{12} - a_{21} a_{12} a_{21} a_{12}  & -2a_{21}+a_{21}a_{12}a_{21}
\end{array}
\right),
\end{eqnarray}
$\bullet$ $p=3$:
\begin{eqnarray}
&&
(\Phi_{\sigma_1^7}^L)_{11}=
4 a_{21} -10 a_{21} a_{12} a_{21} +6 a_{21} a_{12} a_{21} a_{12} a_{21}
-a_{21} a_{12} a_{21} a_{12} a_{21} a_{12} a_{21},
\nonumber \\
&&
(\Phi_{\sigma_1^7}^L)_{12}=
 1-6a_{21}a_{12} +5 a_{21} a_{12} a_{21} a_{12}
-a_{21}a_{12}a_{21}a_{12}a_{21}a_{12},
\nonumber \\
&&
(\Phi_{\sigma_1^7}^L)_{21}=
-1+6a_{12}a_{21} - 5a_{12} a_{21} a_{12} a_{21}
+a_{12}a_{21}a_{12}a_{21} a_{12}a_{21},
\nonumber \\
&&
(\Phi_{\sigma_1^7}^L)_{22}=
 3a_{12}-4a_{12}a_{21}a_{12} +a_{12}a_{21} a_{12}a_{21}a_{12},
\\
&&
(\Phi_{\sigma_1^7}^R)_{11}=
4 a_{12} -10 a_{12} a_{21} a_{12} +6 a_{12} a_{21} a_{12} a_{21} a_{12}
-a_{21} a_{21} a_{12} a_{21} a_{12} a_{21} a_{12},
\nonumber \\
&&
(\Phi_{\sigma_1^7}^L)_{12}=
 -1+6a_{12}a_{21} +5 a_{12} a_{21} a_{12} a_{21}
+a_{12}a_{21}a_{12}a_{21}a_{12}a_{21},
\nonumber \\
&&
(\Phi_{\sigma_1^7}^L)_{21}=
1-6a_{21}a_{12} - 5 a_{21} a_{12} a_{21} a_{12}
+a_{21}a_{12}a_{21}a_{12}a_{21} a_{12}
\nonumber \\
&&
(\Phi_{\sigma_1^7}^L)_{22}=
 3a_{21}-4a_{21}a_{12}a_{21} +a_{21}a_{12}a_{21} a_{12}a_{21}.
\end{eqnarray}

For the writhe $w(\sigma^{2p+1})=2p+1$ and self-linking number ${\rm
sl}(\sigma^{2p+1})=2p-1$,\footnote{
The self-linking number $sl(B)$ is defined by
\begin{eqnarray}
\text{sl}(B)=w(B)-n(B),
\end{eqnarray}
where $n(B)$ denotes the number of braid strands. For $(2,2p+1)$-torus
knots $n(B)=2$.

The authors greatly appreciate to Lenhard Ng for explanation on this
choice of self-linking number, and pointing out how to correct
table \ref{table_aug},  (\ref{super2aug}), and (\ref{aug2AV}) by a
choice of $\text{sl}(B)=1$.
}
we can describe the degree-0 infinity transverse homology
$HT_0^{\infty}(T^{(2,2p+1)})$ of $(2,2p+1)$ torus knot
combinatorially.
To find the augmentation polynomial
${\rm Aug}_{T^{(2,2p+1)}}(\mu,\lambda;U,V)$, we treat $a_{ij}$ as an abelian
variable. The condition (\ref{DGA_B}) relates $a_{12}$ and $a_{21}$
\begin{eqnarray}
a_{21}=\frac{\mu^{2p+1}U^p}{\lambda V^p}a_{12},
\end{eqnarray}
and we denote $a_{12}=:x$ in the following.

With the above prerequisites, the ideals ${\mathcal I}^{(2,2p+1)}$ for
$$
HT_0^{\infty}(T^{(2,2p+1)})\simeq
(\mathbb{Z}[\lambda^{\pm 1},\mu^{\pm 1},U^{\pm 1},V^{\pm 1}])[x]/{\mathcal I}^{(2,2p+1)},
$$
which consist of polynomials $f_s^{(2,2p+1)}(\mu,\lambda,U,V;x)$, (for $s=1,2$),
can now be found by appropriate manipulations in the ideal ${\cal I}_{K}$ in the matrix form:
${\cal I}_{K}=\bigl\{\hat{\bf A}-{\bf \Lambda}_B'\cdot \Phi_B^L\cdot \check{\bf A},\;
\check{\bf A}-\cdot \hat{\bf A}\cdot\Phi_B^R\cdot ({\bf \Lambda}_B')^{-1}\bigr\}$. We find the following results for several low values of $p$:\\
$\bullet$ $p=1$:
\begin{eqnarray}
 f_1^{(2,3)}&=&V^2 \lambda + V \lambda \mu - U x
 \mu^3 - U x^2 \mu^4, \\
 f_2^{(2,3)}&=&V \lambda + U V \lambda \mu - V x \lambda \mu -
 U x^2 \mu^3.
\nonumber
\end{eqnarray}
$\bullet$ $p=2$:
\begin{eqnarray}
 f_1^{(2,5)}&=&
V^5 \lambda^2 + V^4 \lambda^2 \mu +
 U^2 V^2 x \lambda \mu^5 - U^2 V^3 x^2 \lambda \mu^5 -
 3 U^2 V^2 x^2 \lambda \mu^6 + U^4 x^4 \mu^{11},
 \\
 f_2^{(2,5)}&=&
V^2 \lambda + V^3 x \lambda + U V^2 \lambda \mu +
 2 V^2 x \lambda \mu - U^2 x^2 \mu^5 - U^2 x^3 \mu^6.
\nonumber
\end{eqnarray}
$\bullet$ $p=3$:
\begin{eqnarray}
 f_1^{(2,7)}&=&
V^{10} \lambda^3 + V^9 \lambda^3 \mu -
 U^3 V^6 x \lambda^2 \mu^7 - 3 U^3 V^7 x^2 \lambda^2 \mu^7 -
 6 U^3 V^6 x^2 \lambda^2 \mu^8
\nonumber \\
&&
+ U^6 V^4 x^4 \lambda \mu^{14} +
 5 U^6 V^3 x^4 \lambda \mu^{15} - U^9 x^6 \mu^{22},
\\
 f_2^{(2,7)}&=&
-V^6 \lambda^2 + 2 V^7 x \lambda^2 - U V^6 \lambda^2 \mu +
 3 V^6 x \lambda^2 \mu + U^3 V^3 x^2 \lambda \mu^7 -
 U^3 V^4 x^3 \lambda \mu^7
\nonumber \\
&&
- 4 U^3 V^3 x^3 \lambda \mu^8 +
 U^6 x^5 \mu^{15}.
\nonumber
\end{eqnarray}

Finally the augmentation polynomials can be found as the resultants of
$f_1^{(2,2p+1)}$ and $f_2^{(2,2p+1)}$ listed above, with respect to the variable $x$.
Explicit augmentation polynomials for $p=1,2,3$, which we obtain
in this way, are listed in table \ref{table_aug}.\footnote{In this table, we omitted some extra
factors which appear in the resultant in the same manner as in super-$A$-polynomials.}
These augmentation polynomials contain a lot of interesting information, and relate to
various incarnations of $A$-polynomials studied in knot theory context.
Firstly, they are related to $t=-1$ specialization
of super-$A$-polynomials. In particular, setting $V=1$ and performing
a change of variables (\ref{super2aug}), we reproduce the results listed in table \ref{table_super},
with precise identification given in (\ref{super2augBis}).
Secondly, after the variable change (\ref{aug2AV}) these augmentation polynomials
reproduce $Q$-deformed $A$-polynomials derived in \cite{AVqdef},
see (\ref{aug2AVbis}) for detailed identification.
Also note that combining the changes of variables (\ref{super2aug}) and (\ref{aug2AV})
gives a direct relation (\ref{super2av_torus}) between $t=-1$ specialization of the super-$A$-polynomial and $Q$-deformed $A$-polynomials of \cite{AVqdef}. Finally, upon the specialization $U=V=1$, the augmentation polynomial
$\text{Aug}_K(\mu,\lambda;U,V)$ reduces \cite{NgFramed} to the $SL(2)$ $A$-polynomial $A_K(\mu,\lambda)$ studied in
\cite{CCGL}.

In view of all the relations discussed above, we expect that the full super-$A$-polynomial
should also be found from an appropriate $t$-deformation of DGA.

\begin{table}[h]
\centering
\begin{tabular}{|@{$\Bigm|$}c|@{$\Bigm|$}l|}
\hline
\textbf{Knot}  & $Aug_K (\mu,\lambda;U,V)$  \\
\hline
\hline
$T^{(2,3)}$ &
{\scriptsize $
 V^2 (V + \mu) \lambda^2
- U V
(V^2 + V \mu + 2 \mu^2 - 2 U V \mu^2 + U \mu^3 + U^2 \mu^4)
\lambda
+U^2 \mu^3 (1 + U \mu) $}
\\
\hline
$T^{(2,5)}$ &
{\scriptsize $V^6(V+\mu)\lambda^3
$}\\
&{\scriptsize $
-U^2 V^4 (V^4 + V^3 \mu + 2 V^2 \mu^2 - 2 U V^3 \mu^2 +
   2 V \mu^3 - 2 U V^2 \mu^3 + 3 \mu^4 - 4 U V \mu^4 +
   U^2 V^2 \mu^4 + U \mu^5
$}\\
&{\scriptsize $\quad
+ U^2 V \mu^5 + 2 U^2 \mu^6)\lambda^2
$}\\
&{\scriptsize $
+U^4 V^2 \mu^5 (2 V^2 + V \mu + U V^2 \mu + 3 \mu^2 -
   4 U V \mu^2 + U^2 V^2 \mu^2 + 2 U \mu^3 - 2 U^2 V \mu^3 +
   2 U^2 \mu^4 - 2 U^3 V \mu^4
$}\\
&{\scriptsize $\quad
+ U^3 \mu^5 + U^4 \mu^6)\lambda
$}\\
&{\scriptsize $
-U^6 \mu^{10} (1 + U \mu)
$}
\\
\hline
$T^{(2,7)}$ &
{\scriptsize $V^{12}  (V + \mu) \lambda^4
$}\\
&{\scriptsize $
-U^3 V^9 (V^6 + V^5 \mu + 2 V^4 \mu^2 - 2 U V^5 \mu^2 +
   2 V^3 \mu^3 - 2 U V^4 \mu^3 + 3 V^2 \mu^4 - 4 U V^3 \mu^4 +
    U^2 V^4 \mu^4 + 3 V \mu^5
$}\\
&{\scriptsize $\quad
- 4 U V^2 \mu^5 +
   U^2 V^3 \mu^5 + 4 \mu^6 - 6 U V \mu^6 + 2 U^2 V^2 \mu^6 +
   U \mu^7 + 2 U^2 V \mu^7 + 3 U^2 \mu^8)\lambda^3
$}\\
&{\scriptsize $
+U^6 V^6 \mu^7 (3 V^4 + 2 V^3 \mu + U V^4 \mu + 6 V^2 \mu^2 -
   8 U V^3 \mu^2 + 2 U^2 V^4 \mu^2 + 3 V \mu^3 -
   2 U V^2 \mu^3 - U^2 V^3 \mu^3 + 6 \mu^4
$}\\
&{\scriptsize $\quad
- 12 U V \mu^4 +
   10 U^2 V^2 \mu^4 - 4 U^3 V^3 \mu^4 + 3 U \mu^5 -
   2 U^2 V \mu^5 - U^3 V^2 \mu^5 + 6 U^2 \mu^6 -
   8 U^3 V \mu^6 + 2 U^4 V^2 \mu^6
$}\\
&{\scriptsize $\quad
 + 2 U^3 \mu^7 +
   U^4 V \mu^7 + 3 U^4 \mu^8)\lambda^2
$}\\
&{\scriptsize $
-U^9 V^3 \mu^{14} (3 V^2 + V \mu + 2 U V^2 \mu + 4 \mu^2 -
   6 U V \mu^2 + 2 U^2 V^2 \mu^2 + 3 U \mu^3 - 4 U^2 V \mu^3 +
    U^3 V^2 \mu^3 + 3 U^2 \mu^4
$}\\
&{\scriptsize $\quad
- 4 U^3 V \mu^4 +
   U^4 V^2 \mu^4 + 2 U^3 \mu^5 - 2 U^4 V \mu^5 +
   2 U^4 \mu^6 - 2 U^5 V \mu^6 + U^5 \mu^7 + U^6 \mu^8)\lambda
$}\\
&{\scriptsize $
+U^{12} \mu^{21} (1 + U \mu)
$}
\\
\hline
\end{tabular}
\caption{Augmentation polynomials for $(2,2p+1)$ torus knots with $p=1,2,3$. For their explicit relations to $t=-1$ specialization of super-$A$-polynomials, as well as other versions of $A$-polynomials arising in knot theory context, see discussion at the end of section \protect\ref{ssec-aug}.
\label{table_aug} }
\end{table}


\section{Grading conventions and variable changes}   \label{app-grading}

There are many different variables and grading conventions used in recent literature on knot theory, refined topological vertex, and Chern-Simons theory; for example, the following symbols are often used: $q,t,q_1,q_2,{\bf q}, {\bf t}, a, {\bf a}, A, Q$, \emph{etc}. The purpose of this appendix is to summarize transformations and variable changes between those various conventions. We stress that, in general, one has to be particularly careful when different gradings are used -- in such cases, transformations between various conventions amount not only to a simple redefinition of variables (for example, transpositions of Young diagrams which label colored superpolynomials may be involved in such a process). Detailed description of such phenomena has been given in \cite{FGS}. In turn, without repeating this underlying machinery, our task here is merely to provide explicitly variable changes between most often conventions used in literature. Moreover, in the second part of the appendix, we summarize relations between notation used in writing super-$A$-polynomials, augmentation polynomials of \cite{NgFramed}, and $Q$-deformed polynomials of \cite{AVqdef}.

\subsection{Choice of grading and variables}

In the knot theory context, the grading conventions used in this paper are the same as in \cite{FGS,GS}: the quantum parameter entering the commutation relations (\ref{xycomm}) is denoted by $q$, the HOMFLY polynomial is expressed in terms of $a$ and $q$, and the Poincar{\'e} polynomial of knot homologies depends on a new parameter $t$, so that the (colored) superpolynomials depend on variables $a,q,t$. Moreover, the unrefined limit (reducing Poincar{\'e} polynomial to the Euler characteristic) is obtained from $t\to -1$ limit (with other parameters fixed), and the reduction from the HOMFLY polynomial to $sl(N)$ quantum group invariant arises upon the substitution $a=q^N$ (with other parameters fixed). In what follows we relate this convention to other choices made in literature.


Firstly, we recall that certain expressions for colored superpolynomials,
such as those for the unknot (\ref{Punknot}) and torus knots
(\ref{Paqt-torus}) considered in this paper, were derived in \cite{FGS}
from the viewpoint of Chern-Simons theory written in
terms of parameters $A,q_1,q_2$.
Here we define the normalized refined Chern-Simons amplitude
\begin{eqnarray}
Z_{\text{ref}}^{R}(K;A,q_1,q_2):=Z_{SU(N)}^{\text{ref}}(\S^3,K_{R};q_1,q_2) / Z_{SU(N)}^{\text{ref}}(\S^3,\unknot_{R};q_1,q_2).
\label{ref_cs_amp}
\end{eqnarray}
For $K=T^{(2,2p+1)}$ with $R=S^r$ and $R=\Lambda^r$,
$Z_{\text{ref}}^{R}(K;A,q_1,q_2)$ yields
\begin{eqnarray}
R = S^r ~:\; &&
Z^{\text{ref}}_{S^r}(T^{(2,2p+1)};A,q_1,q_2)
\nonumber \\
&&  = \;
A^{r/2}q_2^{-r/2}\frac{(q_2;q_1)_r}{(A;q_1)_r}
\sum_{\ell=0}^r
\frac{(q_2;q_1)_{\ell}(q_1;q_1)_r(A;q_1)_{r+\ell}(q_2^{-1}A;q_1)_{r-\ell}}
{(q_1;q_1)_{\ell}(q_2;q_1)_r(q_2;q_1)_{r+\ell}(q_1;q_1)_{r-\ell}}\frac{(1-q_2q_1^{2\ell})}{(1-q_2q_1^{r+\ell})}
\nonumber \\
&&\quad \quad\quad\quad \quad\quad\quad\quad\quad\quad\;\times
A^{-r}
q_1^{\frac{r-\ell}{2}}
q_2^{\frac{3r-\ell}{2}}
\left[
(-1)^{r-\ell}A^{\frac{r}{2}}q_1^{\frac{r^2-\ell^2}{2}}q_2^{-\frac{\ell}{2}}\right]^{2p+1} \,,
\label{Braid_symmetric}
\\
R = \Lambda^r ~:\; && Z^{\text{ref}}_{\Lambda^r}(T^{(2,2p+1)};A,q_1,q_2)
\nonumber \\
&& = \;
(-1)^rA^{-r/2}q_2^{-r/2}\frac{(q_2;q_2)_r}{(A^{-1};q_2)_r}
\sum_{\ell=0}^r\frac{(q_1;q_2)_{\ell}(q_2;q_2)_{r+\ell}(A^{-1};q_2)_{r+\ell}(q_1^{-1}A^{-1};q_2)_{r-\ell}}
{(q_2;q_2)_{\ell}(q_1q_2;q_2)_{r+\ell}(q_2;q_2)_{r+\ell}(q_2;q_2)_{r-\ell}}
\nonumber \\
&&\quad\quad\quad\quad  \quad\quad\quad\quad\quad\quad\;\times
\frac{(1-q_1q_2^{2\ell})}{(1-q_1)}
A^{r}q_1^{r-\frac{\ell}{2}}q_2^{r-\frac{\ell}{2}}
\left[(-1)^{\ell}A^{\frac{r}{2}}q_1^{\frac{\ell}{2}}q_2^{\frac{\ell^2-r^2}{2}}
\right]^{2p+1},
\label{Braid_anti-symmetric}
\end{eqnarray}
where $A=q_2^N$.

In table \ref{change}, we describe the change of variables between 
the colored superpolynomial $\P^{S^r}(K;a,q,t)$ and the refined
Chern-Simons amplitude $Z_{\text{ref}}^{S^r}(K;A,q_1,q_2)$ for the GS
grading.
This change of variable can be found by 
combining (\ref{GSvarchng}) and further change of variables 
\begin{eqnarray}
(A,q_1,q_2)\mapsto (A,q_2^{-1},q_1^{-1})
\end{eqnarray}
which exchanges the refined Chern-Simons amplitudes (\ref{ref_cs_amp})
between $R=S^r$ and $R=\Lambda^r$.
For $(2,2p+1)$ torus knots, 
$Z^{\text{ref}}_{S^r/\Lambda^r}(T^{(2,2p+1)};A,q_1,q_2)$ 
coincides with the colored superpolynomial
$\P^{S^r/\Lambda^r}(T^{(2,2p+1)};a,q,t)$ by implementing a factor 
and the change of variables in table \ref{change}
\begin{eqnarray}
\P^{S^r}(T^{(2,2p+1)};a,q,t)&=&(-1)^{pr}\left(\frac{q_1}{q_2}\right)^{pr/2}Z_{\text{ref}}^{S^r}(T^{(2,2p+1)};A,q_1,q_2), \\
\P^{\Lambda^r}(T^{(2,2p+1)};a,q,t)&=&(-1)^{pr}\left(\frac{q_2}{q_1}\right)^{pr/2}Z_{\text{ref}}^{\Lambda^r}(T^{(2,2p+1)};A,q_1,q_2).
\end{eqnarray}

On the other hand, the colored superpolynomial for figure-eight knot is derived in appendix \ref{app-figure8} using variables $A_I,q_I,t_I$ (which should be identified respectively with $A,q,t$ in \cite{ItoyamaMMM}). In order to transform these variables to our grading and $a,q,t$ parameters in which the colored superpolynomial (\ref{Paqt41}) is written, we need the following transformation
\be
q_I^2 = q,\qquad  A_I^2 = - a t^3,\qquad t_I = - t q^{1/2}   \label{AIqItI-aqt}.
\ee
This transformation can be found as a composition of transformations given in the second and the third row of table \ref{change}. 
More generally, for other pairs of conventions, table \ref{change} can be used to find
how they are related, by reducing them to our $a,q,t$ choice in an intermediate step.

\begin{table}[h]
\centering
\begin{tabular}{|@{$\Bigm|$}c|@{$\Bigm|$}c|@{$\Bigm|$}c|}
\hline
Variables  & Change of variables & Inverse relation\\
\hline
\hline
Refined $SU(N)$ CS \cite{AS} & $(t^N,q,t)$ &
$(A,q_1,q_2)$
\\
$(t^N,q,t)$ & $=(A,q_1,q_2)$ &$=(t^N,q,t)$ \\
\hline
Refined CS in \cite{DMMSS,ItoyamaMMM} 
& $(A_I,q_I,t_I)$ & $(A,q_1,q_2)$ \\
$(A_I,q_I,t_I)$ 
& $=(A^{1/2},q_1^{1/2},-q_2^{1/2})$& $=(A_I^2,q_I^2,t_I^2)$ \\
\hline
GS grading: $\P^{S^r}$ \cite{GS} & $(a,q,t)$ & $(A,q_1,q_2)$\\
$(a,q,t)$ & $=\left(A(q_1/q_2)^{3/2},q_1,-(q_2/q_1)^{1/2}\right)$
     & $=(-at^3,q,qt^{2})$ \\
\hline
GS grading: $\P^{\Lambda^r}$ \cite{GS} & $(a,q,t)$ & $(A,q_1,q_2)$\\
$(a,q,t)$ & $=\left(A(q_2/q_1)^{1/2},q_2,-(q_1/q_2)^{1/2}\right)$ & $=(-at,qt^2,q)$\\
\hline
DGR grading \cite{DGR}& $(a,q,t)$ & $(A,q_1,q_2)$\\
$(a,q,t)$ & $=\left(A^{1/2}(q_2/q_1)^{1/4},q_2^{1/2},-(q_1/q_2)^{1/2}\right)$ & $=(-a^2t,q^2t^2,q^{2})$\\
\hline
HOMFLY in \cite{AVqdef} & $(Q,q)$ & $(a,q)$ with $t=-1$ \\
$(Q,q)$ & $=(a,q)$ with $t=-1$ & $=(Q,q)$ \\
\hline
\end{tabular}
\caption{Changes of variables between various conventions used in literature. \label{change} }
\end{table}

\subsection{Conventions for (deformed) $A$-polynomials and augmentation polynomials}  \label{ssec-Aaug}

Upon substitution $t=-1$ the super-$A$-polynomial reduces to $a$-deformed version of the $A$-polynomial, which already appeared in literature under two different guises: as the so-called augmentation polynomial of knot contact homology in \cite{NgFramed}, and as $Q$-deformed $A$-polynomial in \cite{AVqdef}. While these three objects are clearly closely related, they were introduced naturally from different perspectives and using different conventions. Here we discuss how to relate these different conventions to each other.

Let us also recapitulate that, firstly, $t=-1$ specialization of super-$A$-polynomial gives a polynomial in variables $x$ and $y$, which depends on a parameter $a$. Secondly, the augmentation polynomial of \cite{NgFramed} is written in terms of variables $\lambda$ and $\mu$, and it depends additionally on $U$ and $V$. Thirdly, the $Q$-deformed $A$-polynomial in \cite{AVqdef} is a polynomial in $\alpha$ and $\beta$, which depends on a parameter $Q$. One should also be aware that explicit relations between these sets of variables may depend on a particular knot which is studied. In what follows we carefully discuss the form of these relations for $(2,2p+1)$ torus knots. Independently, the identification of variables between super-$A$-polynomial and $Q$-deformed $A$-polynomial for the figure-eight knots is given in (\ref{super2av_fig8}).

Let us focus now on the case of $(2,2p+1)$ torus knots, and relate first
$t=-1$ specialization of super-$A$-polynomial
$A^{\textrm{super}}(x,y;a,-1)$ to the augmentation polynomial
$\text{Aug}(\mu,\lambda;U,1)$. The relation between our variables
$(x,y)$ and the variables $(\mu,\lambda)$ can be easily deduced from the specialization to ordinary $A$-polynomial, {\it cf.} \eqref{AugA}.
Indeed, as proved in \cite[Proposition 5.9]{NgFramed},
the augmentation polynomial contains $A$-polynomial as a factor.
For instance, the ordinary $A$-polynomial for the trefoil knot is given in \eqref{AAAtref},
and comparing with the results in \cite{Ng}
we see that the following identification must be made
\begin{eqnarray}
&&A^{\text{super}} (x, y; a=1,t=-1) = -(x-1)(y-1)(y+x^3)
\nonumber \\
&=&- \text{Aug}(\mu,\lambda;U=1,V=1)
= (\lambda - 1)
(\mu + 1) (\lambda - \mu^3)  .  \nonumber
\end{eqnarray}
In consequence we deduce the following relation between the variables:
\be
x  = - \mu, \qquad y  = \lambda.  \label{xymulambda}
\ee


More generally, we would like to understand the role of our parameter $a$ from the contact homology viewpoint.
We find that for $(2,2p+1)$ torus knots, if one applies the following identification\footnote{This change of variables holds for $p=1,2,3$ at least.}
\be
x=-\mu,\quad y=\frac{1+\mu}{1+U\mu}\lambda,\quad  t=-1, \quad a=U, \quad V=1,
\label{super2aug}
\ee
then the super-$A$-polynomial and the augmentation polynomial are related as follows
\be
A_{T^{(2,2p+1)}}^{\text{super}}(x,y;a,t=-1)  =\frac{(1+\mu)^p}{(1+U\mu)^{p+1}}
\text{Aug}_{T^{(2,2p+1)}}(\mu,\lambda;U,V=1).
\label{super2augBis}
\ee
Note that for $U=1$ the relations (\ref{super2aug}) reduce to (\ref{xymulambda}).

On the other hand, one can relate the augmentation polynomial for torus knots
to the $Q$-deformed $A$-polynomial in \cite{AVqdef} by the following identification of parameters
\be
\lambda=Q^p\beta^{4p+2}\alpha,\quad \mu=-\beta,\quad U=Q,  \label{aug2AV}
\ee
which leads to the relation
\be
\text{Aug}_{T^{(2,2p+1)}}(\mu,\lambda;U,V=1) =  \beta^{p(2p+1)}Q^{p+1}A^{\textrm{Q-def}}_{T^{(2,2p+1)}}(\alpha,\beta;Q).
\label{aug2AVbis}
\ee

Combining transformations (\ref{super2aug}) and (\ref{aug2AV}), one can directly relate super-$A$-polynomial to $Q$-deformed $A$-polynomial. The resulting identification of parameters is given explicitly in (\ref{super2av_torus}).

\bigskip


Finally let us explain why in (\ref{super2av_fig8}), (\ref{super2av_torus}), as well as (\ref{super2aug}) a meromorphic factor $\frac{1-Q \beta}{1-\beta}=\frac{1-ax}{1-x}$ appears in the change of variable $y$.
This factor is a consequence of the fact that the super-$A$-polynomial discussed in this paper arises from analysis of reduced colored superpolynomials, while the augmentation polynomial or $Q$-deformed polynomial considered in \cite{AVqdef,NgFramed} are related to unreduced knot invariants.
The reduced superpolynomial differs from the unreduced one by the
unknot factor
\begin{eqnarray}
\bar{\P}^{R}(K;a,q,t)=\bar{\P}^{R}(\unknot;a,q,t)\P^{R}(K;a,q,t).
\end{eqnarray}
One explicit example of such a relation is given in (\ref{Pour-Kawagoe}) for figure-eight knot (where the overall sign difference arises from a particular choice of unknot normalization, and this is not relevant in what follows). For the symmetric representation $R=S^{r=n-1}$, the unreduced colored superpolynomial for the unknot $\bar{\P}_n(\unknot;a,q,t)$
is given explicitly in (\ref{Punknot}), and obeys the
recursion relation (\ref{Punrec}).

Now the recursion relation for the reduced superpolynomial
\begin{eqnarray}
\hat{A}^{\text{super}}(\hat{x},\hat{y};a,q,t)\P_n=0
\end{eqnarray}
leads to the following recursion for the unreduced superpolynomial
\begin{eqnarray}
0&=&\left[\bar{\P_n}(\unknot)
\hat{A}^{\text{super}}(\hat{x},\hat{y};a,q,t)\bar{\P_n}(\unknot)^{-1}
\right]\bar{\P_n}
=\hat{A}^{\text{super}}(\hat{x'},\hat{y'};a,q,t)\bar{\P_n}
\nonumber \\
&=&:\hat{A'}^{\text{super}}(\hat{x},\hat{y};a,q,t)\bar{\P_n}.
\end{eqnarray}
where
\begin{eqnarray}
\hat{x'}=\bar{\P_n}(\unknot)\hat{x}\bar{\P_n}(\unknot)^{-1},\quad
\hat{y'}=\bar{\P_n}(\unknot)\hat{y}\bar{\P_n}(\unknot)^{-1}.
\end{eqnarray}
Using the recursion relation (\ref{Punrec}), we find
\begin{eqnarray}
&& \hat{x'}=\bar{\P_n}(\unknot) \bar{\P_n}(\unknot)^{-1}\hat{x} = \hat{x}, \\
&&
\hat{y'}
=\bar{\P}_{n}(\unknot)\bar{\P}_{n+1}(\unknot)^{-1}\hat{y}
=(-at^{3}q)^{1/2}\frac{1-q^{n-1}}{1+at^3q^{n-1}}\hat{y}.
\end{eqnarray}
Therefore we conclude that in the classical limit $\hbar\to 0$, the super-A-polynomial for
reduced and unreduced superpolynomials, $A^{\text{super}}(x,y;a,t)$ and
$A^{\prime\;\text{super}}(x',y';a,t)$, are related by the change of variables
\begin{eqnarray}
x=x',\quad y=(-at^{3})^{-1/2}\frac{1+at^3x}{1-x}y'.
\end{eqnarray}
For $t=-1$ this change of variables further reduces to
\begin{eqnarray}
x=x',\quad y=a^{-1/2}\frac{1-ax}{1-x}y'.
\end{eqnarray}
This is therefore the origin of the meromorphic factor $\frac{1-ax}{1-x}$ in the
change of $y$-variables in (\ref{super2av_fig8}), (\ref{super2av_torus}) and (\ref{super2aug}).


\section{Super-$A$-polynomial for $x=1$, and for $(2,9)$ and $(2,11)$ torus knot}   \label{app-T211}
\begin{table}[H]
\centering
\begin{tabular}{|@{$\Bigm|$}c|@{$\Bigm|$}l|}
\hline
\textbf{Knot}  & $A_K^{\text{super}} (x,y;a,t)$  \\
\hline
\hline
${\bf 4_1}$ & $y^3-a^{-1}t^{-2}(1+at(1+t+t^2(1+at)))y^2$ \\
\hline
$T^{(2,3)}$ & $y^2-a(1+t^2(1+at))y$ \\
\hline
$T^{(2,5)}$ & $y^3-a^2(1+t^2(1+at)(1+t^2))y^2$ \\
\hline
$T^{(2,7)}$ & $y^4-a^3(1+t^2(1+at)(1+t^2+t^4))y^2$ \\
\hline
$T^{(2,9)}$ & $y^4-a^4(1+t^2(1+at)(1+t^2+t^4+t^6))y^2$ \\
\hline
$T^{(2,11)}$ & $y^4-a^5(1+t^2(1+at)(1+t^2+t^4+t^6+t^8))y^2$ \\
\hline
\end{tabular}
\caption{Super-$A$-polynomials with $x=1$ for the ${\bf 4_1}$ knot and $(2,2p+1)$ torus knot up to $p=5$.}
\label{superAlimit}
\end{table}

\begin{table}[H]
\centering
\begin{tabular}{|@{$\Bigm|$}c|@{$\Bigm|$}l|}
\hline
\textbf{Knot}  & $A_K^{\text{super}} (x,y;a,t)$  \\
\hline
\hline
$T^{(2,9)}$
 & {\scriptsize $y^5
-$}
$\frac{a^4}{1 +a t^3 x} $
{\scriptsize $
(1 - t^2 x + 2 t^2 x^2 + 2 a t^3 x^2 - 2 t^4 x^3 - 2 a t^5 x^3 +
  3 t^4 x^4 + 4 a t^5 x^4 + a^2 t^6 x^4 - 3 t^6 x^5 - 4 a t^7 x^5
$}\\
&{\scriptsize $\quad\quad
-
  a^2 t^8 x^5 + 4 t^6 x^6 + 6 a t^7 x^6 + 2 a^2 t^8 x^6 - 4 t^8 x^7 -
  6 a t^9 x^7 - 2 a^2 t^{10} x^7 + 5 t^8 x^8 + 8 a t^9 x^8 +
  3 a^2 t^{10} x^8
$}\\
&{\scriptsize $\quad\quad
+ a t^{11} x^9 - 3 a^2 t^{12} x^9 + 4 a^2 t^{12} x^{10})y^4
$}
\\
&{\scriptsize $+$}$\frac{a^8 t^{10} (-1 + x) x^9 }{(1 +a t^3 x)^2}$
{\scriptsize
$(4 - 3 t^2 x + a t^3 x + 9 t^2 x^2 + 12 a t^3 x^2 + 3 a^2 t^4 x^2 -
  6 t^4 x^3 - 6 a t^5 x^3 + 12 t^4 x^4 + 24 a t^5 x^4
$}
\\
&{\scriptsize $\quad
+
  18 a^2 t^6 x^4 + 6 a^3 t^7 x^4 - 6 t^6 x^5 - 9 a t^7 x^5 -
  6 a^2 t^8 x^5 - 3 a^3 t^9 x^5 + 10 t^6 x^6 + 24 a t^7 x^6 +
  27 a^2 t^8 x^6 + 16 a^3 t^9 x^6
$}
\\
&{\scriptsize $\quad
+ 3 a^4 t^{10} x^6 + 4 a t^9 x^7 -
  6 a^3 t^{11} x^7 - 2 a^4 t^{12} x^7 + 12 a^2 t^{10} x^8 +
  18 a^3 t^{11} x^8 + 6 a^4 t^{12} x^8 + 3 a^3 t^{13} x^9 -
  3 a^4 t^{14} x^9
$}
\\
&{\scriptsize $\quad
+ 6 a^4 t^{14} x^{10})y^3
$}
\\
&{\scriptsize $-$}$\frac{a^{12} t^{20} (-1 + x)^2 x^{18} }{(1 +a t^3 x)^3}$
{\scriptsize $(6 - 3 t^2 x + 3 a t^3 x + 12 t^2 x^2 + 18 a t^3 x^2 +
  6 a^2 t^4 x^2 - 4 t^4 x^3 + 6 a^2 t^6 x^3 + 2 a^3 t^7 x^3
$}
\\
&{\scriptsize $\quad
+
  10 t^4 x^4 + 24 a t^5 x^4 + 27 a^2 t^6 x^4 + 16 a^3 t^7 x^4 +
  3 a^4 t^8 x^4 + 6 a t^7 x^5 + 9 a^2 t^8 x^5 + 6 a^3 t^9 x^5 +
  3 a^4 t^{10} x^5
$}
\\
&{\scriptsize $\quad
+ 12 a^2 t^8 x^6 + 24 a^3 t^9 x^6 +
  18 a^4 t^{10} x^6 + 6 a^5 t^{11} x^6 + 6 a^3 t^{11} x^7 +
  6 a^4 t^{12} x^7 + 9 a^4 t^{12} x^8 + 12 a^5 t^{13} x^8 +
  3 a^6 t^{14} x^8
$}
\\
&{\scriptsize $\quad
+ 3 a^5 t^{15} x^9 - a^6 t^{16} x^9 + 4 a^6 t^{16} x^{10})y^2
$}\\
&{\scriptsize $+$}$\frac{a^{16} t^{30} (-1 + x)^3 x^{27} }{(1 +a t^3 x)^4}$
{\scriptsize
$(4 - t^2 x + 3 a t^3 x + 5 t^2 x^2 + 8 a t^3 x^2 + 3 a^2 t^4 x^2 +
 4 a t^5 x^3 + 6 a^2 t^6 x^3 + 2 a^3 t^7 x^3
$}
\\
&{\scriptsize $\quad
+ 4 a^2 t^6 x^4 +
 6 a^3 t^7 x^4 + 2 a^4 t^8 x^4 + 3 a^3 t^9 x^5 + 4 a^4 t^{10} x^5 +
 a^5 t^{11} x^5 + 3 a^4 t^{10} x^6 + 4 a^5 t^{11} x^6 + a^6 t^{12} x^6
$}
\\
&{\scriptsize $\quad
+
 2 a^5 t^{13} x^7 + 2 a^6 t^{14} x^7 + 2 a^6 t^{14} x^8 + 2 a^7 t^{15} x^8 +
 a^7 t^{17} x^9 + a^8 t^{18} x^{10})y
$}\\
&{\scriptsize $-$}$\frac{a^{20} t^{40} (-1 + x)^4 x^{36} }{(1 +a t^3 x)^4}$
{\scriptsize
$
$}\\
\hline
\end{tabular}
\caption{Super-$A$-polynomial for $(2,2p+1)$ torus knot with $p=4$. For
 super-$A$-polynomials for torus knots with $p=1,2,3,5$ see tables
 \protect\ref{table_super} and \protect\ref{table_super5}.
\label{table_super4} }
\end{table}

\begin{table}[H]
\centering
\begin{tabular}{|@{$\Bigm|$}c|@{$\Bigm|$}l|}
\hline
\textbf{Knot}  & $A_K^{\text{super}} (x,y;a,t)$  \\
\hline
\hline
$T^{(2,11)}$ &
{\tiny $y^6
-\frac{a^5}{1 +a t^3 x}
(1 - t^2 x + 2 t^2 x^2 + 2 a t^3 x^2 - 2 t^4 x^3 - 2 a t^5 x^3 +
  3 t^4 x^4 + 4 a t^5 x^4 + a^2 t^6 x^4 - 3 t^6 x^5 - 4 a t^7 x^5
-
  a^2 t^8 x^5
$}\\
&{\tiny $\quad\quad
+ 4 t^6 x^6 + 6 a t^7 x^6 + 2 a^2 t^8 x^6 - 4 t^8 x^7 -
  6 a t^9 x^7 - 2 a^2 t^{10} x^7 + 5 t^8 x^8 + 8 a t^9 x^8 +
  3 a^2 t^{10} x^8 - 5 t^{10} x^9 - 8 a t^{11} x^9
$}\\
&{\tiny $\quad\quad
- 3 a^2 t^{12} x^9 +
  6 t^{10} x^{10} + 10 a t^{11} x^{10} + 4 a^2 t^{12} x^{10} + a t^{13} x^{11} -
  4 a^2 t^{14} x^{11} + 5 a^2 t^{14} x^{12})y^5
$}\\
&{\tiny $+\frac{a^{10}t^{12} (-1 + x) x^{11} }{(1 +a t^3 x)^2}
(5 - 4 t^2 x + a t^3 x + 12 t^2 x^2 + 16 a t^3 x^2 + 4 a^2 t^4 x^2 -
  9 t^4 x^3 - 10 a t^5 x^3 - a^2 t^6 x^3 + 18 t^4 x^4 +
  36 a t^5 x^4
$}\\
&{\tiny $\quad
+ 26 a^2 t^6 x^4 + 8 a^3 t^7 x^4 - 12 t^6 x^5 -
  21 a t^7 x^5 - 14 a^2 t^8 x^5 - 5 a^3 t^9 x^5 + 20 t^6 x^6 +
  48 a t^7 x^6 + 48 a^2 t^8 x^6 + 24 a^3 t^9 x^6
$}\\
&{\tiny $\quad
+ 4 a^4 t^{10} x^6 -
  10 t^8 x^7 - 20 a t^9 x^7 - 21 a^2 t^{10} x^7 - 14 a^3 t^{11} x^7 -
  3 a^4 t^{12} x^7 + 15 t^8 x^8 + 40 a t^9 x^8 + 52 a^2 t^{10} x^8 +
  36 a^3 t^{11} x^8
$}\\
&{\tiny $\quad
+ 9 a^4 t^{12} x^8 + 5 a t^{11} x^9 - 4 a^2 t^{12} x^9 -
  15 a^3 t^{13} x^9 - 6 a^4 t^{14} x^9 + 20 a^2 t^{12} x^{10} +
  32 a^3 t^{13} x^{10} + 12 a^4 t^{14} x^{10} + 4 a^3 t^{15} x^{11}
$}\\
&{\tiny $\quad
-
  6 a^4 t^{16} x^{11} + 10 a^4 t^{16} x^{12})y^4
$}\\
&{\tiny $-\frac{a^{15}t^{24} (-1 + x)^2 x^{22} }{(1 +a t^3 x)^3}
(10 - 6 t^2 x + 4 a t^3 x + 24 t^2 x^2 + 36 a t^3 x^2 +
  12 a^2 t^4 x^2 - 12 t^4 x^3 - 9 a t^5 x^3 + 6 a^2 t^6 x^3 +
  3 a^3 t^7 x^3
$}\\
&{\tiny $\quad
+ 30 t^4 x^4 + 72 a t^5 x^4 + 72 a^2 t^6 x^4 +
  36 a^3 t^7 x^4 + 6 a^4 t^8 x^4 - 10 t^6 x^5 - 12 a t^7 x^5 -
  3 a^2 t^8 x^5 + 2 a^3 t^9 x^5 + 3 a^4 t^{10} x^5 + 20 t^6 x^6
$}\\
&{\tiny $\quad
+
  60 a t^7 x^6 + 96 a^2 t^8 x^6 + 92 a^3 t^9 x^6 + 48 a^4 t^{10} x^6 +
  12 a^5 t^{11} x^6 + 10 a t^9 x^7 + 12 a^2 t^{10} x^7 + 3 a^3 t^{11} x^7 -
  2 a^4 t^{12} x^7 - 3 a^5 t^{13} x^7
$}\\
&{\tiny $\quad
+ 30 a^2 t^{10} x^8 +
  72 a^3 t^{11} x^8 + 72 a^4 t^{12} x^8 + 36 a^5 t^{13} x^8 +
  6 a^6 t^{14} x^8 + 12 a^3 t^{13} x^9 + 9 a^4 t^{14} x^9 -
  6 a^5 t^{15} x^9 - 3 a^6 t^{16} x^9
$}\\
&{\tiny $\quad
+ 24 a^4 t^{14} x^{10} +
  36 a^5 t^{15} x^{10} + 12 a^6 t^{16} x^{10} + 6 a^5 t^{17} x^{11} -
  4 a^6 t^{18} x^{11} + 10 a^6 t^{18} x^{12})y^3
$}\\
&{\tiny $+\frac{a^{20}t^{36} (-1 + x)^3 x^{33} }{(1 +a t^3 x)^4}
(10 - 4 t^2 x + 6 a t^3 x + 20 t^2 x^2 + 32 a t^3 x^2 +
  12 a^2 t^4 x^2 - 5 t^4 x^3 + 4 a t^5 x^3
$}\\
&{\tiny $\quad
+ 15 a^2 t^6 x^3 +
  6 a^3 t^7 x^3 + 15 t^4 x^4 + 40 a t^5 x^4 + 52 a^2 t^6 x^4 +
  36 a^3 t^7 x^4 + 9 a^4 t^8 x^4 + 10 a t^7 x^5 + 20 a^2 t^8 x^5 +
  21 a^3 t^9 x^5
$}\\
&{\tiny $\quad
+ 14 a^4 t^{10} x^5 + 3 a^5 t^{11} x^5 +
  20 a^2 t^8 x^6 + 48 a^3 t^9 x^6 + 48 a^4 t^{10} x^6 +
  24 a^5 t^{11} x^6 + 4 a^6 t^{12} x^6 + 12 a^3 t^{11} x^7 +
  21 a^4 t^{12} x^7
$}\\
&{\tiny $\quad
+ 14 a^5 t^{13} x^7 + 5 a^6 t^{14} x^7 +
  18 a^4 t^{12} x^8 + 36 a^5 t^{13} x^8 + 26 a^6 t^{14} x^8 +
  8 a^7 t^{15} x^8 + 9 a^5 t^{15} x^9 + 10 a^6 t^{16} x^9 + a^7 t^{17}
     x^9
$}\\
&{\tiny $\quad
+
  12 a^6 t^{16} x^{10} + 16 a^7 t^{17} x^{10} + 4 a^8 t^{18} x^{10} +
  4 a^7 t^{19} x^{11} - a^8 t^{20} x^{11} + 5 a^8 t^{20} x^{12})y^2
$}\\
&{\tiny $-\frac{a^{25}t^{48} (-1 + x)^4 x^{44} }{(1 +a t^3 x)^5}
(5 - t^2 x + 4 a t^3 x + 6 t^2 x^2 + 10 a t^3 x^2 + 4 a^2 t^4 x^2 +
  5 a t^5 x^3 + 8 a^2 t^6 x^3 + 3 a^3 t^7 x^3 + 5 a^2 t^6 x^4 +
  8 a^3 t^7 x^4
$}\\
&{\tiny $\quad
+ 3 a^4 t^8 x^4 + 4 a^3 t^9 x^5 + 6 a^4 t^{10} x^5 +
  2 a^5 t^{11} x^5 + 4 a^4 t^{10} x^6 + 6 a^5 t^{11} x^6 + 2 a^6 t^{12} x^6 +
  3 a^5 t^{13} x^7 + 4 a^6 t^{14} x^7 + a^7 t^{15} x^7
$}\\
&{\tiny $\quad
+ 3 a^6 t^{14} x^8 +
  4 a^7 t^{15} x^8 + a^8 t^{16} x^8 + 2 a^7 t^{17} x^9 + 2 a^8 t^{18} x^9 +
  2 a^8 t^{18} x^{10} + 2 a^9 t^{19} x^{10} + a^9 t^{21} x^{11} + a^{10} t^{22} x^{12})y
$}\\
&{\tiny $+\frac{a^{30}t^{60} (-1 + x)^5 x^{55} }{(1 +a t^3 x)^5}$}
\\
\hline
\end{tabular}
\caption{Super-$A$-polynomial for $(2,2p+1)$ torus knot with $p=5$. For
 super-$A$-polynomials for torus knots with $p=1,2,3,4$ see tables
 \protect\ref{table_super} and \protect\ref{table_super4}.
\label{table_super5} }
\end{table}


\newpage

\bibliographystyle{JHEP_TD}
\bibliography{abmodel}

\end{document}